\documentclass[letterpaper,twocolumn,twoside]{ieeetran}
\usepackage{graphics} 
\usepackage{epsfig} 
\usepackage{amsmath} 
\usepackage{amssymb}  
\usepackage{cite}
\usepackage{color}
\usepackage{graphicx}
\usepackage{algpseudocode}
\usepackage{algorithm}
\usepackage{tabularx}
\usepackage[font=footnotesize]{caption}

\renewcommand{\theenumi}{(\alph{enumi})}

\newtheorem{theorem}{Theorem}[section]
\newtheorem{definition}[theorem]{Definition}
\newtheorem{assumption}[theorem]{Assumption}
\newtheorem{lemma}[theorem]{Lemma}

\newtheorem{remark}[theorem]{Remark}
\newtheorem{example}[theorem]{Example}
\newtheorem{corollary}[theorem]{Corollary}
\newtheorem{proposition}[theorem]{Proposition}

\newcommand{\naturals}{\mathbb{N}}
\newcommand{\real}{\mathbb{R}}
\newcommand{\cplx}{\mathbb{C}}

\newcommand{\range}{\mathcal{R}}

\newcommand{\Dc}{\mathcal{D}}

\newcommand{\Fc}{\mathcal{F}}

\newcommand{\Kc}{\mathcal{K}}
\newcommand{\Lc}{\mathcal{L}}
\newcommand{\Mc}{\mathcal{M}}

\newcommand{\Sc}{\mathcal{S}}

\newcommand{\EDMD}[3]{\operatorname{EDMD}(#1,#2,#3)}
\newcommand{\until}[1]{\{1,\dots,#1\}}
\newcommand{\rank}[1]{\operatorname{rank}(#1)}
\newcommand{\map}[3]{#1:#2 \rightarrow #3}

\newcommand{\dx}{D(X)}
\newcommand{\dy}{D(Y)}
\newcommand{\dsx}{D_S^X}
\newcommand{\dsy}{D_S^Y}

\newcommand{\tdx}{\tilde{D}(X)}
\newcommand{\tdy}{\tilde{D}(Y)}
\newcommand{\cdx}{\Dc(X)}
\newcommand{\cdy}{\Dc(Y)}
\newcommand{\Dtilde}{\tilde{D}_{\operatorname{aprx}}}
\newcommand{\Span}{\operatorname{span}}

\newcommand{\re}{\operatorname{Re}}
\newcommand{\im}{\operatorname{Im}}

\newcommand{\rows}{{\operatorname{rows}}}
\newcommand{\cols}{{\operatorname{cols}}}
\newcommand{\rown}{{\operatorname{\sharp rows}}}
\newcommand{\coln}{{\operatorname{\sharp cols}}}
\newcommand{\basis}{{\operatorname{basis}}}
\newcommand{\ssd}{{\operatorname{SSD}}}

\newcommand{\cssd}{C^{\ssd}}
\newcommand{\cssdapprox}{C^{\ssd}_{\operatorname{aprx}}}
\newcommand{\csinf}{\cssd_\infty}

\newcommand{\Kssd}{K^{\ssd}}
\newcommand{\Kssdapprox}{K^{\ssd}_{\operatorname{aprx}}}
\newcommand{\Ksinf}{\Kssd_\infty}

\newcommand{\new}[1]{{\color{blue} #1}}
\renewcommand{\new}[1]{#1}

\newcommand{\longthmtitle}[1]{\mbox{}{\textit{(#1):}}}

\newcommand{\oprocendsymbol}{\hbox{$\square$}}
\newcommand{\oprocend}{\relax\ifmmode\else\unskip\hfill\fi\oprocendsymbol}

\renewcommand{\baselinestretch}{.998}

\allowdisplaybreaks

\title{
  Learning Koopman Eigenfunctions and Invariant Subspaces from Data:
  Symmetric Subspace Decomposition
  \thanks{This work was supported by ONR Award N00014-18-1-2828.}
  \thanks{A preliminary version of this work appeared
    as~\cite{MH-JC:19-cdc} at the IEEE Conference on Decision and
    Control.}}

\author{Masih Haseli and Jorge Cort\'es
  \thanks{Masih Haseli and Jorge Cort\'es are with Department of
    Mechanical and Aerospace Engineering, University of California,
    San Diego, CA 92093, USA, {\tt\small
      \{mhaseli,cortes\}@ucsd.edu}}
}

\graphicspath{{epsfiles/}}

\begin{document}	
	
\maketitle

\begin{abstract}
  This paper develops data-driven methods to identify eigenfunctions
  of the Koopman operator associated to a dynamical system and
  subspaces that are invariant under the operator.  We build on
  Extended Dynamic Mode Decomposition (EDMD), a data-driven method
  that finds a finite-dimensional approximation of the Koopman
  operator on the span of a predefined dictionary of functions.  We
  propose a necessary and sufficient condition to identify Koopman
  eigenfunctions based on the application of EDMD forward and backward
  in time.  
  Moreover, we propose the Symmetric Subspace Decomposition (SSD)
  algorithm, an iterative method which provably identifies the
  maximal Koopman-invariant subspace and the Koopman eigenfunctions in
  the span of the dictionary.  We also introduce the Streaming
  Symmetric Subspace Decomposition (SSSD) algorithm, an online
  extension of SSD that only requires a small, fixed memory and
  incorporates new data as is received.
  Finally, we propose an extension of SSD that approximates
    Koopman eigenfunctions and invariant subspaces when the dictionary
    does not contain sufficient informative eigenfunctions.
\end{abstract}

\section{Introduction}
Driven by advances in processing, data storage, cloud services, and
algorithms, the world has witnessed in recent years a revolution in
data-driven learning, analysis, and control of dynamical
phenomena. State-space, probabilistic, and neural network models are
among the most popular methods to model dynamical systems. With
sufficient a priori information about the dynamics, state-space
methods can provide closed-form analytic models that describe
accurately the dynamical behavior. Such models, however, are generally
nonlinear and their analytical study becomes arduous for moderate to
high-dimensional systems. Probabilistic approaches, on the other hand,
provide an alternative description that is conductive to dealing with
incomplete information about the underlying dynamics. However, under
such approaches, deriving mathematical guarantees may be hard, if not
impossible. Neural networks can describe the dynamics with high
accuracy given enough data. The models acquired by neural networks are
highly nonlinear and difficult to study analytically. Hence, even
though they can be successful in predicting the behavior of the
system, they often do not provide a deeper understanding into their
dynamics. These reasons have motivated researchers to seek alternative
strategies to capture the dynamics using data with minimum a priori
information in a computationally efficient way that result in simple
yet accurate models. Approximating the Koopman operator associated
with a dynamical system is one of such strategies. The Koopman
operator is a linear but generally infinite-dimensional operator that
fully describes the behavior of the underlying dynamical system. Even
though the linearity of the Koopman operator makes its spectral
properties a powerful tool for analysis, its infinite-dimensional
nature prevents the use of conventional linear algebraic tools
developed to work with digital computers.  One way to circumvent this
issue is to identify finite-dimensional subspaces that are invariant
under the Koopman operator.  This paper develops data-driven methods
to identify such subspaces.

\textit{Literature Review:} The Koopman
operator~\cite{BOK:31,BOK-JVN:32} is a linear but generally
infinite-dimensional operator that provides an alternative view of
dynamical systems by describing the effect of the dynamics on a
functional space. Being a linear operator enables one to use its
spectral properties to capture and predict the behavior of nonlinear
dynamical systems~\cite{IM:05,CWR-IM-SB-PS-DSH:09,MB-RM-IM:12}. This
leads to a wide variety of applications including state
estimation~\cite{AS-MOW-MM-AB:17,MN-LM:18}, system
identification~\cite{AM-JG:20,DB-CDR-RV:19,BK-PG-PB-SN-MB:17},
sensor and actuator placement~\cite{SS-UV-RR:16}, model
reduction~\cite{SK-FN-PK-HW-IK-CS-FN,AA-JNK:17},
control~\cite{MK-IM-automatica:18,HA-MK-IM:18,SP-SK:17,BH-XM-UV:18,EK-JNK-SLB:17,AS-DE:17,AN-JSK:20},
and robotics~\cite{GM-MC-XT-TM:19,MLC-AH-GM-TG-TM-XT:20}.
Moreover, the
eigenfunctions of the Koopman operator play an important role in
stability analysis of nonlinear systems~\cite{AM-IM:16}. Due to the
infinite-dimensional nature of the Koopman operator, the digital
implementation of the aforementioned applications is not possible
unless one can find a way to represent the effect of the operator on
finite-dimensional subspaces.  The literature has explored several
data-driven methods to find such finite-dimensional approximations,
which can be divided into two main categories: projection methods and
invariant-subspace methods. Projection methods fit a linear model to
the data acquired from the system. The most popular approach in this
category is Dynamic Mode Decomposition (DMD), first proposed to
capture dynamical information from fluid flows~\cite{PJS:10}. DMD uses
linear algebraic methods 
to form a linear model from time series data. The
work~\cite{KKC-JHT-CWS:12} explores the properties of DMD and its
connection with the Koopman operator,
and~\cite{JHT-CWR-DML-SLB-JNK:13} generalizes it to work with
non-sequential data snapshots.  Several extensions perform online computations to work with streaming datasets~\cite{MSH-MOW-CWR:14,HZ-CWR-EAD-LNC:19,SA-KM:19}, account for the effect of measurement noise on data~\cite{STMD-MSH-MOW-CWR:16,MSH-CWR-EAD-LNC:17}, promote
sparsity~\cite{MRJ-PJS-JWN:14}, and consider time-lagged data
snapshots~\cite{SLC-JMV:17}. Extended Dynamic Mode Decomposition
(EDMD)~\cite{MOW-IGK-CWR:15} is an important variations of DMD that
lifts the states of the system to a (generally higher-dimensional)
functional space using a predefined dictionary of functions and finds
the projection of the Koopman operator on that subspace. The
work~\cite{MK-IM:18} studies the convergence properties of EDMD to the
Koopman operator as the number of data snapshots and dictionary
elements go to infinity.
EDMD is specifically designed to work with exact data and experiments
and simulations show that it may not work well with noisy data. Our
previous work~\cite{MH-JC:19-acc} presented a noise-resilient
extension of EDMD able to work with data corrupted with measurement noise. The basic lifting idea of EDMD can also be combined with known information
about the dynamics  to increase the accuracy of the
model~\cite{EQ-BK-BP-KW:20}. The aforementioned methods provide linear
higher-dimensional approximations for the underlying dynamics that
are, however, not suitable for long term predictions, since they are
generally not exact. This issue can be tackled by finding subspaces
that are invariant under the Koopman operator, since the acquired
linear models are exact over them. This is the subject of the second
group of approaches. 
The works~\cite{QL-FD-EMB-IGK:17,NT-YK-TY:17,EY-SK-NH:19,SEO-CWR:19} provide
approaches to find functions that span Koopman-invariant subspaces
using neural networks. Moreover, since Koopman eigenfunctions span
Koopman-invariant subspaces, one can use the empirical methods
provided in~\cite{SLB-BWB-JLP-JNK:16,EK-JNK-SLB:17} to approximate the
Koopman eigenfunctions and consequently the invariant
subspaces. Moreover, the work in~\cite{MK-IM:20} provides theoretical
and empirical results based on multi-step predictions to find Koopman
eigenfunctions. Interestingly, note that none of the aforementioned
methods provide mathematical guarantees for the identified functions
to be Koopman eigenfunctions.

\textit{Statement of Contributions:} We present data-driven methods to
identify Koopman eigenfunctions and Koopman-invariant subspaces
associated with a potentially nonlinear dynamical system. First, we
study the properties of the standard EDMD method regarding the
identification of Koopman eigenfunctions. We prove that EDMD correctly
identifies all the Koopman eigenfunctions in the span of the
predefined dictionary. This necessary condition however is not
sufficient, i.e., the functions identified by the EDMD method are not
necessarily Koopman eigenfunctions. This motivates our next
contribution, which is a necessary and sufficient condition that
characterizes the functions that evolve linearly according to the
available data snapshots. This condition is based on the application
of EDMD forward and backward in time. The identified functions are not
necessarily Koopman eigenfunctions, since one can only guarantee that
they evolve linearly on the available data (but not necessarily
starting anywhere in the state space). However, we prove that under
reasonable assumptions on the density of the sampling, the identified
functions are Koopman eigenfunctions almost surely. Our next
contribution seeks to provide computationally efficient ways of
identifying Koopman eigenfunctions and Koopman-invariant subspaces.
In fact, checking the aforementioned necessary and sufficient
condition requires one to calculate and compare the eigendecomposition
of two potentially large matrices, which can be computationally
cumbersome. Moreover, even though the subspace spanned by all the
eigenfunctions in the span of the original dictionary is
Koopman-invariant, it might not be maximal. To address these
limitations, we propose the Symmetric Subspace Decomposition (SSD)
strategy, which is an iterative method to find the maximal subspace
that remains invariant under the application of dynamics (and its
associated Koopman operator) according to the available data. We prove
that SSD also finds all the functions that evolve linearly in time
according to the available data. Moreover, we prove that under the
same conditions in the sampling density, the SSD strategy identifies
the maximal Koopman-invariant subspace in the span of the original
dictionary almost surely.
Our next contribution is motivated by applications where the data
becomes available in an online fashion. In such scenarios, at any
given time step, one would need to perform SSD on all the available
data received up to that time.  Performing SSD requires the
calculation of several singular value decompositions for matrices that
scale with the size of the data, in turn requiring significant memory
capabilities. To address these shortcomings, we propose the Streaming
Symmetric Subspace Decomposition (SSSD) strategy, which refines the
calculated Koopman-invariant subspaces each time it receives new data
and deals with matrices of fixed and relatively small size
(independent of the size of the data). We prove that SSSD and SSD
methods are equivalent, in the sense that for a given dataset, they
both identify the same maximal Koopman-invariant subspace.  Our
  last contribution is motivated by the fact that, in some cases the
  predefined dictionary does not contain sufficient eigenfunctions to
  capture important information from the dynamics. To address this
  issue, we provide an extension of SSD, termed Approximated-SSD,
  enabling us to approximate Koopman eigenfunctions and invariant
  subspaces. We show how the accuracy of the approximation can be
  tuned using a design parameter.


\section{Preliminaries}\label{sec:preliminaries}
In this section\footnote{We denote by $\naturals$, $\naturals_0$,
  $\real$, $\real_{\geq 0}$, and $\cplx$, the sets of natural,
  nonnegative integer, real, positive real, and complex numbers
  respectively. For a matrix $A \in \cplx^{m \times n}$, we denote the
  sets comprised of its rows by $\rows(A)$, its columns by $\cols(A)$,
  the number of its rows by $\rown(A)$, and the number of its columns
  by $\coln(A)$, respectively.  In addition, we denote its
  pseudo-inverse, transpose, complex conjugate, conjugate transpose,
  Frobenius norm, and range space by $A^\dagger$, $A^T$, $\bar{A}$,
  $A^H$, $\|A\|_F$, and $\range(A)$, respectively.  For $1\leq i < k
  \leq m$, we denote by $A_{i:k}$ the matrix formed with the $i$th to
  $k$th rows of $A$. Moreover, $A_{i,j}$ denotes the $ij$th
    element of $A$. For a square nonsingular matrix $B$, we denote
  its inverse by $B^{-1}$.  Given matrices $A \in \cplx^{m \times n}$
  and $B \in \cplx ^{m \times d}$, we denote by $[A,B] \in \cplx^{m
    \times (n+d)}$ the matrix created by concatenating $A$ and
  $B$. The angle between vectors $v, w \in \real^n$ is
  $\angle(v,w)$. Given $v_1,\ldots, v_k \in \cplx^n$,
  $\Span\{v_1,\ldots, v_k \}$ represents the set comprised of all
  linear combinations $c_1v_1+\cdots+c_nv_n$, with $c_1,\dots, c_n \in
  \cplx$. We use $j$ to denote the imaginary unit (the solution of
  $x^2+1=0$). For $v \in \cplx^n$, we denote its real and imaginary
  parts by $\re(v)$ and $\im(v)$, and its 2-norm as $\|v\|_2 :=
  \sqrt{v^H v}$. Given a set $A$, we denote its complement by
  $A^c$. Given sets $A$ and $B$, $A \subseteq B$ means that $A$ is a
  subset of $B$. We denote by $A \cap B$ and $A \cup B$ the
  intersection and union of $A$ and $B$, and set $A \setminus B := A
  \cap B^c$. Given a sequence of sets $\{A_i\}_{i=1}^{\infty}$, we
  denote its superior and inferior limits by $\limsup_{i \to \infty}
  A_i$ and $\liminf_{i \to \infty} A_i$, respectively. We refer to the
  set consisting of all continuous strictly increasing functions
  $\alpha : \real_{\geq 0} \to \real_{\geq 0}$ with $\alpha(0)=0$ by
  class-$\Kc$. Given $\map{f}{B}{A}$ and $\map{g}{C}{B}$, $\map{f
    \circ g}{C}{A}$ denotes their composition.}, we review basic
concepts on the Koopman operator and Extended Dynamic Mode
Decomposition.

\subsection{Koopman Operator}\label{sec:Koopman}
Here, we introduce the (discrete-time) Koopman operator and its
spectral properties following~\cite{MB-RM-IM:12}. Consider a
nonlinear, time-invariant, continuous map $T:\Mc \to \Mc$ on $\Mc
\subseteq \real^n$, defining the dynamical system
\begin{align}\label{eq:dymamical-sys}
  x^+ = T(x).
\end{align}
The dynamics~\eqref{eq:dymamical-sys} acts on the points in the state
space $\Mc$ and generates trajectories of the system. The Koopman
operator, on the other hand, provides an alternative approach to
analyze~\eqref{eq:dymamical-sys} based on evolution of functions (also
known as observables) defined on $\Mc$ and taking values in
$\cplx$. Formally, let $\Fc$ be a linear space of functions from $\Mc$
to $\cplx$ which is closed under composition with $T$, i.e.,
\begin{align}\label{eq:Fc-closedness}
  f \circ T \in \Fc, \quad \forall f \in \Fc .
\end{align}
The Koopman operator $\Kc: \Fc \to \Fc$ associated
with~\eqref{eq:dymamical-sys} is
\begin{align*}
  \Kc(f) = f \circ T.
\end{align*}
A closer look at the definition of the Koopman operator shows that it
advances the observables in time, i.e., for $g = \Kc(f)$ then
\begin{align}\label{eq:observable-advance-time}
  g(x) = f \circ T (x)= f(x^+), \quad \forall x \in \Mc.
\end{align}
This equation shows how the Koopman operator encodes the dynamics on
the functional space $\Fc$. The operator is linear as a direct
consequence of linearity in $\Fc$, i.e., for every $f_1, f_2 \in \Fc$
and $c_1,c_2 \in \cplx$,
\begin{align}\label{eq:Koopman-spatial-linear}
  \Kc (c_1 f_1+ c_2 f_2) = c_1 \Kc(f_1) + c_2 \Kc(f_2).
\end{align}
Assuming $\Fc$ contains the functions describing the states of the
system, $g_i(x)=x_i$ with $i \in \until{n}$, the Koopman operator
fully characterizes the global features of the dynamics in a linear
fashion. Moreover, the operator might be (and generally is) infinite
dimensional either by choice of $\Fc$ or due to closedness requirement
in~\eqref{eq:Fc-closedness}.

Being linear, one can naturally define its eigendecomposition. A
function $\phi \in \Fc$ is an \emph{eigenfunction} of $\Kc$ associated
with \emph{eigenvalue} $\lambda \in \cplx$ if
\begin{align}\label{eq:Koopman-eigendecomposition}
  \Kc(\phi) = \lambda \phi.
\end{align}
The combination of~\eqref{eq:observable-advance-time}
and~\eqref{eq:Koopman-eigendecomposition} leads to a significant
property of the Koopman operator: the linear evolution of its
eigenfunctions in time. Formally, given an eigenfunction $\phi$,
\begin{align}\label{eq:eig-temporal-evo}
  \phi(x^+) =(\phi \circ T)(x) = \Kc(\phi)(x) = \lambda \phi(x).
\end{align}
The linear evolution of eigenfunctions, together with
linearity~\eqref{eq:Koopman-spatial-linear}, enables us to use
spectral properties to analyze the nonlinear
system~\eqref{eq:dymamical-sys}. Given a set of eigenpairs $\{(
\lambda_i, \phi_i)\}_{i=1}^{N_k}$ such that $\Kc (\phi_i) = \lambda_i
\phi_i, i \in \until{N_k}$, one can describe the evolution of every
function $f$ in $\Span (\{\phi_i\}_{i=1}^{N_k})$, i.e., $ f =
\sum_{i=1}^{N_k} c_i \phi_i$, for some $\{c_i\}_{i=1}^{N_k} \subset
\cplx$, as
\begin{align}\label{eq:function-evolution-Koopman}
  f(x(k)) = \sum_{i=1}^{N_k} c_i \lambda_i^k \, \phi_i (x(0)), \quad
  \forall k \in \naturals_0.
\end{align}
The constants $\{c_i\}_{i=1}^{N_k}$ are called \emph{Koopman
  modes}. It is important to note that one might need to use $N_k =
\infty$ to fully describe the behavior of the dynamical system.

Another important notion in the analysis of the Koopman operator is
the invariance of subspaces under its application. Formally, a
subspace $\Sc \subseteq \Fc$ is \emph{Koopman-invariant} if for every
$f \in \Sc$ we have $\Kc(f) \in \Sc$. Furthermore, $\Sc$ is
\emph{maximal Koopman-invariant} in $\Lc \subseteq \Fc$ if it contains every
Koopman-invariant subspace in $\Lc$. Naturally, a set comprised of
Koopman eigenfunctions spans a Koopman-invariant subspace.

\subsection{Extended Dynamic Mode Decomposition}\label{sec:EDMD}
Our exposition here mainly follows~\cite{MOW-IGK-CWR:15}. As mentioned
earlier, despite its linearly, the infinite-dimensional nature of the
Koopman operator obstructs the use of efficient linear algebraic
methods. One natural way to overcome this problem is finding
finite-dimensional approximations for it. Extended Dynamic Mode
Decomposition (EDMD) is a popular data-driven method to perform this
task that lifts data snapshots acquired from the dynamical system to a
higher-dimensional space using a predefined dictionary of functions.
The projection of the action of the operator on the span of the
dictionary can then be found by solving a least-squares problem.

Formally, let $\Dc: \real^n \to \real^{1 \times N_d}$ be a dictionary of
$N_d$ functions in $\Fc$ with $ \Dc(x)=[d_1(x), \ldots
,d_{N_d}(x)]$. Moreover, let $X,Y \in \real^{N \times n}$ be matrices
comprised of $N$ data snapshots such that $y_i = T(x_i)$ for $i \in
\until{N}$, where $x_i^T$ and $y_i^T$ are $i$th rows of $X$ and $Y$,
respectively. For convenience, we define the action of the dictionary
on a matrix as
\begin{align*}
  \cdx := [\Dc(x_1)^T, \dots , \Dc(x_N)^T]^T.
\end{align*}
The EDMD method approximates the projection of the Koopman operator by
finding the matrix that best explains the data over the dictionary, i.e.,
\begin{align*}
  \underset{K}{\text{minimize}} \| \cdy - \cdx K \|_F^2,
\end{align*}
which yields the closed-form solution
\new{
\begin{align}\label{eq:EDMD-closed-form}
  K^*= \EDMD{\Dc}{X}{Y} := \cdx^\dagger \cdy.
\end{align}
} Note that the solution depends on the choice of dictionary. If the
dictionary spans a Koopman-invariant subspace, then $\| \Dc(Y) -
\Dc(X) K^* \|_F^2 = 0$ and $K^*$ fully captures the evolution of
functions in $\Span(\Dc(x))$. Otherwise, EDMD loses some information
about the dynamics.

\section{Problem Statement}\label{sec:problem-statement}
As described in Section~\ref{sec:EDMD}, the EDMD method loses
information about the dynamical system when the employed dictionary
does not span a Koopman-invariant subspace. As a result, in such
cases, the EDMD approximation is not appropriate for long term
prediction of the state evolution.  Motivated by this observation, our
goal is to find the maximal Koopman-invariant subspace and Koopman
eigenfunctions in the span of a given dictionary. 

Formally, given the dynamical system~\eqref{eq:dymamical-sys} defined
by $T:\Mc \to \Mc$, data matrices $X$ and $Y$ comprised of $N$ data
snapshots, and an \new{arbitrary} dictionary of functions $D$, our main goal is
two-fold:
\begin{enumerate}
\item find all the Koopman eigenfunctions in $\Span(D(x))$;
\item find a basis for the maximal Koopman-invariant subspace in
  $\Span(D(x))$.
\end{enumerate} 
Note that (a) and (b) are closely related. The eigenfunctions found by
solving~(a) span Koopman-invariant subspaces. Those invariant
subspaces however might not be maximal. This mild difference between
(a) and (b) requires the use of different solution approaches.  Since we are
dealing with finite-dimensional linear subspaces, we aim to use linear
algebra instead of optimization-based methods, which are widely used
for solving these types of problems. This enables us to directly use
computationally efficient linear algebraic packages that optimization
methods rely on.

Throughout the paper, we use the following assumption regarding the
dictionary snapshots.

\begin{assumption}\longthmtitle{Full Column Rank Dictionary
    Matrices}\label{a:full-rank}
  The matrices $\dx$ and $\dy$ have full column rank.  \oprocend
\end{assumption}

Assumption~\ref{a:full-rank} is reasonable: in order to hold, the
dictionary functions must be linearly independent, i.e., the functions
must form a basis for $\Span(D(x))$. Moreover, the assumption requires
the set of initial conditions $\rows(X)$ to be diverse enough to
capture important characteristics of the dynamics.  Our
treatment here relies on EDMD, which is not specifically designed to
work with data corrupted with measurement noise. Hence, we
assume access to data with high signal-to-noise ratio. In practice,
one might need to pre-process the data to use the algorithms proposed
here.

\section{EDMD and Koopman Eigenfunctions}\label{sec:EDMD-charac}
Here we investigate the capabilities and limitations of the EDMD
method regarding the identification of Koopman
eigenfunctions.
\new{Throughout the paper, we use the following notations to represent the EDMD matrices applied on data matrices $X$ and $Y$ forward and backward in time
\begin{align*}
K_f & = \EDMD{D}{X}{Y} , \quad
K_b  = \EDMD{D}{Y}{X}.    
\end{align*}}
The next result shows that EDMD is not only able to
capture Koopman eigenfunctions but also all the functions that evolve
linearly according to the available data.

\begin{lemma}\longthmtitle{EDMD Captures the Koopman Eigenfunctions in
    the Span of the Dictionary}\label{l:EDMD-captures-eig} 
  Suppose Assumption~\ref{a:full-rank} holds. Let $f(x) = D(x)v$ for
  some $v \in \cplx^{N_d}\setminus \{0\}$ and all $x \in \Mc$.
  \begin{enumerate}
  \item Let $f$ evolve linearly according to the available data, i.e.,
    there exists $\lambda \in \cplx$ such that $f(y_i) = \lambda
    f(x_i)$ for every $i \in \until{\rown(X)}$.
    Then, the vector $v$ is an eigenvector of $K_f$ with eigenvalue
    $\lambda$;
  \item Let $f$ be an eigenfunction of the Koopman operator with
    eigenvalue $\lambda$. Then, the vector $v$ is an eigenvector of
    $K_f$ with eigenvalue $\lambda$.
  \end{enumerate} 
\end{lemma}
\begin{proof}
  (a) Based on the linear evolution of $f$, we have $ \dy v= \lambda
  \dx v$.  Moreover, using the closed-form solution of EDMD, we have $
  K_f \,v = \dx ^\dagger \dy v = \lambda \dx ^\dagger \dx v =
  \lambda v$, where the last equality follows from
  Assumption~\ref{a:full-rank}.
  
  (b) Based on the definition of Koopman eigenfunction, we have
  $f(x^+) = \lambda f(x)$. Since this linear evolution reflects in
  data snapshots, we have $f(y_i) = \lambda f(x_i)$ for every $i \in
  \until{\rown(X)}$ where $x_i^T$ and $y_i^T$ are the $i$th rows of
  $X$ and $Y$ respectively. The rest follows from~(a).
\end{proof}

Despite its simplicity, this result provides significant insight into
the EDMD method. Lemma~\ref{l:EDMD-captures-eig} shows that EDMD can
capture eigenfunctions in the span of the dictionary even if the
underlying subspace is not Koopman invariant. In the literature, it is
well known that the (E)DMD method can capture physical constraints,
conservation laws, and other properties of the underlying system,
which actually correspond to Koopman eigenfunctions, e.g.,
see~\cite{MOW-IGK-CWR:15,SK-FN-SP-JHN-CC-CS:19}. We note that
Lemma~\ref{l:EDMD-captures-eig} 
is a generalization of~\cite[Theorem~1]{JHT-CWR-DML-SLB-JNK:13} to
EDMD when the underlying system is not necessarily linear (or cannot
be approximated by a linear system accurately) and the underlying
subspace is not Koopman invariant.  \new{The next result shows that
  EDMD accurately predicts the evolution of functions in the span of
  Koopman eigenfunctions evaluated on the system's trajectories.

  \begin{proposition}\longthmtitle{EDMD Accurately Predicts Evolution
      of any Linear Combination of Eigenfunctions on System's
      Trajectories}\label{p:EDMD-prediction}
    Let $f(x) = D(x)v$ for some $v \in \cplx^{N_d}\setminus \{0\}$ and
    all $x \in \Mc$.  Assume $f$ is in the span of eigenfunctions
    $\{\phi_i\}_{i=1}^{m} \subset \Span(D)$ with corresponding
    eigenvalues $\{\lambda_i\}_{i=1}^{m} \subset \cplx$. Then, given
    any trajectory $\{x(j)\}_{j=0}^\infty$
    of~\eqref{eq:dymamical-sys},
    \begin{align}\label{eq:long-term-pred-EDMD}
      f(x(j)) = D(x(0)) K_f^j v, \; \forall j \in \naturals_0.
    \end{align}
  \end{proposition}
 \begin{proof}
   Since $f \in \Span(\{\phi_i\}_{i=1}^{m})$, there exist scalars
   $\{c_i\}_{i=1}^{m} \subset \cplx$ such that
   \begin{align}\label{eq:f-in-eig-span}
     f(x) = \sum_{i=1}^m c_i \phi_i(x).
   \end{align}
   Since $\{\phi_i\}_{i=1}^{m} \subset \Span(D)$, there exist vectors
   $\{w_i\}_{i=1}^{m} \subset \cplx^{N_d}$ such that
   \begin{align}\label{eq:eig-vector-form}
     \phi_i(x) = D(x) w_i, \; \forall i \in \until{m}.
   \end{align}
   Combining~\eqref{eq:f-in-eig-span} and~\eqref{eq:eig-vector-form}
   with the definition of $f$, we deduce that $\sum_{i=1}^m c_i w_i =
   v$. Now,
   \begin{align*}
     D(x(0)) K_f^j v &= D(x(0)) K_f^j \sum_{i=1}^m c_i w_i
     \nonumber \\
     &= D(x(0)) \sum_{i=1}^m c_i \lambda_i^j w_i,
   \end{align*}
   where in the last equality we have used
   Lemma~\ref{l:EDMD-captures-eig}(b) for the eigenfunctions
   $\phi_i$s. Using~\eqref{eq:eig-vector-form},
   \begin{align*}
     D(x(0)) K_f^j v = \sum_{i=1}^m c_i \lambda_i^j \phi_i(x(0)) =
     \sum_{i=1}^m c_i \phi_i(x(j)),
   \end{align*}
   where in the second equality we have used the linear temporal
   evolution of Koopman eigenfunctions~\eqref{eq:eig-temporal-evo}. The
   proof is now complete by noting that the previous equation holds for
   all $j \in \naturals_0$ and its the right-hand side is equal to
   $f(x(j))$ based on~\eqref{eq:f-in-eig-span}.
 \end{proof}
}

Lemma~\ref{l:EDMD-captures-eig} provides a necessary condition for the
identification of Koopman eigenfunctions. This condition however is
not sufficient, see e.g.~\cite[Example~IV.3]{MH-JC:19-cdc} for a
counter example.
Interestingly, if a function evolves linearly forward in time, it also
evolves linearly backward in time. The next result shows that checking
this observation provides a necessary and sufficient condition for
identification of functions that evolve linearly in time according to
the available data.

\begin{theorem}\longthmtitle{Identification of Linear Evolutions by
    Forward and Backward EDMD}\label{t:forward-backward-EDMD-evo} 
  Suppose Assumption~\ref{a:full-rank} holds.  Let $f(x) = D(x)v$
  for some $v \in \cplx^{N_d} \setminus \{0\}$.  Then $f(y_i) =
  \lambda f(x_i)$ for some $\lambda \in \cplx \setminus \{0\}$ and for
  all $i \in \until{\rown(X)}$ if and only if $v$ is an eigenvector of
  $K_f$ with eigenvalue~$\lambda$, and an eigenvector of $K_b $ with
  eigenvalue~$\lambda^{-1}$.
\end{theorem}
\begin{proof}
  $(\Leftarrow)$: Using the closed-form solutions of the EDMD problem
  and Assumption~\ref{a:full-rank}, one can write,
  \begin{subequations}
    \begin{align*}
      K_f &= (\dx^T \dx)^{-1} \dx^T \dy,
      \\
      K_b &= (\dy^T \dy)^{-1} \dy^T \dx.
    \end{align*}
  \end{subequations}
  Using these along with the definition of the eigenpair, 
  \begin{subequations}
    \begin{align}
      \lambda \dx^T \dx v &= \dx^T \dy v, \label{eq:eigen1}
      \\
      \lambda^{-1} \dy^T \dy v &= \dy^T \dx v. \label{eq:eigen2}
    \end{align}
  \end{subequations}
  By multiplying~\eqref{eq:eigen1} from the left by $v^H$ and
  using~\eqref{eq:eigen2},
  \begin{align*}
    \lambda \|\dx v\|_2^2 =v^H \dx^T \dy v 
    = \bar{\lambda}^{-1} \|\dy v \|_2^2
  \end{align*}
  which implies
  \begin{align}\label{eq:norm-adjustment}
    |\lambda|^2 \|\dx v\|_2^2 = \|\dy v\|_2^2 .
  \end{align}
  Now, we decompose $\dy v$ orthogonally as
  \begin{align}\label{eq:dnyv-decomposition}
    \dy v = c \dx v + w,
  \end{align}
  with $v^H \dx^T w=0$. Substituting~\eqref{eq:dnyv-decomposition}
  into~\eqref{eq:eigen1} and multiplying both sides from the left by
  $v^H$ yields
  \begin{align*}
    \lambda v^H \dx^T \dx v = c v^H \dx^T \dx v .
  \end{align*}
  Since $v \neq 0$, and under Assumption~\ref{a:full-rank}, we deduce
  that $c = \lambda$.  Substituting the value of $c$
  in~\eqref{eq:dnyv-decomposition}, finding the 2-norm, and using the
  fact that $v^H \dx^T w=0$, one can write
  \begin{align*}
    \|\dy v\|_2^2 = |\lambda|^2 \|\dx v\|_2^2 + \|w\|_2^2.
  \end{align*}
  Comparing this with~\eqref{eq:norm-adjustment}, one deduces that
  $w=0$ and $\dy v = \lambda \dx v$. The result follows by looking at
  this equality in a row-wise manner and noting that $f(x) = D(x)v$.
  
  $(\Rightarrow)$: Based on Lemma~\ref{l:EDMD-captures-eig}(a), $v$
  must be an eigenvector of $K_f$ with eigenvalue $\lambda$.
  Moreover, since $\lambda \neq 0$ one can write $f(x_i) =
  \lambda^{-1} f(y_i)$ for every $i \in \until{\rown(X)}$ and,
  consequently, using Lemma~\ref{l:EDMD-captures-eig}(a) once again,
  we have $K_b v = \lambda^{-1} v$, concluding the proof.
\end{proof}

If the function $f$ satisfies the conditions provided by
Theorem~\ref{t:forward-backward-EDMD-evo}, then $f(x^+) = \lambda
f(x)$ for all $x \in \rows(X)$.  However,
Theorem~\ref{t:forward-backward-EDMD-evo} does not guarantee that $f$
is an eigenfunction, i.e., there is no guarantee that $f(x^+) =
\lambda f(x)$ for all $x \in \Mc$. To circumvent this issue, we
introduce next infinite sampling and make an assumption about its
density.

\begin{assumption}\longthmtitle{Almost sure dense sampling from a
    compact state space}\label{a:dense-sampling}
  Assume the state space $\Mc$ is compact.  Suppose we gather
  infinitely (countably) many data snapshots.  For $N \in \naturals$,
  the first $N$ data snapshots are represented by matrices $X_{1:N}$
  and $Y_{1:N}$ such that $y_i = T(x_i)$ for all $i \in \until{N}$,
  where $x_i$ and $y_i$ are the $i$th rows of $X_{1:N}$ and $Y_{1:N}$,
  respectively (we refer to the columns of $X_{1:N}^T$ as the set
  $S_N$ of initial conditions).
  Assume there exists a class-$\Kc$ function $\alpha$ and sequence
  $\{p_N\}_{N=1}^\infty \subset [0,1]$ such that, for every $N \in
  \naturals$,
  \begin{align*}
    \forall m \in \Mc, \, \exists x \in S_N \; \operatorname{such\;
      that} \, \|m-x\|_2 \leq \alpha \Big( \frac{1}{N} \Big)
  \end{align*}
  holds with probability $p_N$, and $\lim_{N \to \infty} p_N =1$.
  \oprocend
\end{assumption}

Assumption~\ref{a:dense-sampling} is not restrictive as, in most
practical cases, the state space is compact or the analysis is limited
to a specific bounded region. Moreover, the data is usually available
on a bounded region due the limited range of sensors. Regarding the
sampling density, Assumption~\ref{a:dense-sampling} holds for most
standard random samplings.

Noting that our methods presented later require
Assumption~\ref{a:full-rank} to hold, we provide a definition for
dictionary matrices acquired from infinite sampling.

\begin{definition}\longthmtitle{$R$-rich Sequence of Dictionary
    Snapshots}\label{d:r-rich-dictionary}
  Let $\{ X_{1:N}\}_{N = 1}^\infty$ and $\{ Y_{1:N}\}_{N = 1}^\infty$
  be the sequence of data snapshot matrices acquired from
  system~\eqref{eq:dymamical-sys}. Given the dictionary $D: \Mc \to
  \real^{1 \times N_d}$, we say the sequence of dictionary snapshot
  matrices is \emph{$R$-rich} if $R = \min \{M \in \naturals \,|
  \rank{D(X_{1:M})} = \rank{D(Y_{1:M})} = N_d\}$ exists ($R$ is called
  \emph{richness constant}).
\end{definition}
\smallskip

In Definition~\ref{d:r-rich-dictionary}, if
\begin{align*}
  \{M \in \naturals \,| \rank{D(X_{1:M})} = \rank{D(Y_{1:M})} =
  N_d\} \neq \emptyset
\end{align*}
then based on the well-ordering principle, see e.g.~\cite[Chapter
0]{GBF:99}, the minimum of the set exists and the sequence of the
dictionary snapshot matrices is $R$-rich. Moreover, given an $R$-rich
sequence of dictionary snapshots matrices $D(X_{1:N})$ and
$D(Y_{1:N})$, Assumption~\ref{a:full-rank} holds for every $N \geq R$.

We are now ready to identify the Koopman eigenfunctions in the span of
the dictionary using forward-backward EDMD.

\begin{theorem}\longthmtitle{Identification of Koopman Eigenfunctions
    by Forward and Backward EDMD}\label{t:forward-backward-EDMD-eig}
  Given an infinite sampling, suppose that the sequence of dictionary
  snapshot matrices is $R$-rich.
  Let $K_f^N = \EDMD{D}{X_{1:N}}{Y_{1:N}}$, $K_b^N =
  \EDMD{D}{Y_{1:N}}{X_{1:N}}$. Given $v \in
  \cplx^{N_d}\setminus\{0\}$ and $\lambda \in \cplx \setminus\{0\}$,
  let $f(x)=D(x)v$. Then,
  \begin{enumerate}
  \item If $f$ is an eigenfunction of the Koopman operator with
    eigenvalue $\lambda$, then $K_f^N v = \lambda v$ and $K_b^N v =
    \lambda^{-1} v$ for every $N \geq R$;
  \item Conversely, and assuming the dictionary functions are
    continuous and Assumption~\ref{a:dense-sampling} holds, if $K_f^N
    v = \lambda v$ and $K_b^N v = \lambda^{-1} v$ for every $N \geq
    R$, then $f$ is an eigenfunction of the Koopman operator with
    probability~1.
  \end{enumerate} 
\end{theorem}
\begin{proof}
  (a) Since $f$ is a Koopman eigenfunction, for every $i \in \naturals
  $ we have $f(y_i)=\lambda f(x_i)$. Moreover, for every $N \geq R$,
  $D(X_{1:N})$ and $D(Y_{1:N})$ have full column rank.  Therefore, the
  result follows from Theorem~\ref{t:forward-backward-EDMD-evo}.
    
  (b) Based on Theorem~\ref{t:forward-backward-EDMD-evo}, we 
  deduce that, for every $N \geq R$
  \begin{align}\label{eq:dictionary-correspondence}
    f(y_i) = \lambda f(x_i) v, \; \forall i \in \until{N},
  \end{align}
  where $x_i^T$ and $y_i^T$ are the $i$th rows of $X_{1:N}$ and
  $Y_{1:N}$ respectively.  Now, define $h(x) := f \circ T (x) -
  \lambda f(x)$. The function $h$ is continuous since $f$ is a linear
  combination of continuous functions and $T$ is also continuous.  By
  inspecting $h$ on the data points and
  using~\eqref{eq:dictionary-correspondence} and the fact that $ y_i =
  T(x_i)$, for all $i \in \until{N}$, one can show that $h(x_i) = f
  \circ T (x_i) - \lambda f(x_i) = f(y_i)-\lambda f(x_i) = 0$ for
  every $i \in \until{N}$.  Moreover, note that based on
  Assumption~\ref{a:dense-sampling}, the set $S_\infty =
  \bigcup_{i=1}^{\infty} S_i$ is dense in $\Mc$ with probability 1 and
  $h(x) =0$ for every $x \in S_\infty$. As a result, $h(x) =0$ on
  $\Mc$ with probability 1. This implies that $f \circ T(x) = \lambda
  f(x)$ for every $x \in \Mc$ almost surely. Consequently, we have
  $\Kc(f) = \lambda f$ almost surely, and the result follows.
\end{proof}

We note that the technique of considering the evolution forward and
backward in time has also been used in the literature for other
purposes, e.g., to alleviate the effect of measurement noise on the
data when performing
DMD~\cite{STMD-MSH-MOW-CWR:16,MSH-CWR-EAD-LNC:17}. To our knowledge,
the use of this technique here for the identification of Koopman
eigenfunctions and invariant subspaces is novel.
Moreover, unlike~\cite[Algorithm 1]{MK-IM:19}, the methods proposed
here do not require access to the system's multi-step trajectories.
Theorems~\ref{t:forward-backward-EDMD-evo}
and~\ref{t:forward-backward-EDMD-eig} provide conditions to identify
Koopman eigenfunctions. The identified eigenfunctions then can span
Koopman-invariant subspaces. However, one still needs to compare $N_d$
potentially complex eigenvectors and their corresponding
eigenvalues. This procedure can be impractical for
large~$N_d$. Moreover, since $\Mc \subseteq \real^n$, the
eigenfunctions of the Koopman operator form complex-conjugate
pairs. Such pairs can be fully characterized using their real and
imaginary parts, which allows to use instead real-valued
functions. This motivates the development of algorithms to directly
identify Koopman-invariant subspaces.

\section{Identification of Koopman-Invariant Subspaces via Symmetric
  Subspace Decomposition (SSD)}\label{sec:SSD}

Here we provide an algorithmic method to identify Koopman-invariant
subspaces in the span of a predefined dictionary and later show how it
can be used to find Koopman eigenfunctions.  \new{With the setup of
  Section~\ref{sec:problem-statement},} given the \new{original}
dictionary $D: \Mc \to \real^{1 \times N_d}$ comprised of $N_d$
linearly independent functions, we aim to find a dictionary $\tilde{D}
: \Mc \to \real^{1 \times \tilde{N}_d}$ with $\tilde{N}_d$ linearly
independent functions such that the elements of $\tilde{D}$ span the
maximal Koopman-invariant subspace in $\Span(D)$. Since
$\Span(\tilde{D})$ is invariant, we have $ \range(\tilde{D} \circ T) =
\range(\tilde{D})$. This equality gets reflected in the data, i.e.,
given snapshot matrices $X$ and~$Y$,
\begin{align}\label{eq:range-new-dictionary-ssd}
  \range(\tilde{D}(Y)) = \range(\tilde{D}(X)).
\end{align}
Moreover, since the elements of $\tilde{D}$ are in the span of
$D$, there exists a full column rank matrix $C$ such that
$\tilde{D}(x) = D(x) C$, for all $x \in \Mc$. Thus
from~\eqref{eq:range-new-dictionary-ssd},
\begin{align}\label{eq:range-old-dictionary-ssd}
  \range(D(Y)C) = \range(D(X)C).
\end{align}
Hence, we can reformulate the problem as a purely linear-algebraic
problem consisting of finding the full column rank matrix $C$ with
maximum number of columns such
that~\eqref{eq:range-old-dictionary-ssd} holds. To solve this problem,
we propose the Symmetric Subspace Decomposition (SSD) method.
The SSD algorithm relies on the fact,
  from~\eqref{eq:range-new-dictionary-ssd}, that
\begin{align*}
  \range(\tilde{D}(Y)) = \range(\tilde{D}(X)) \subseteq \range(\dx)
  \cap \range(\dy).
\end{align*}
This fact can alternatively be expressed using the null space of the
concatenation~$[\dx,\dy]$.  SSD uses the null space to prune the
dictionary and remove functions that do not evolve linearly in time
according to the available data to identify a potentially smaller
dictionary. At each iteration, SSD repeats the aforementioned
procedure of (i)~concatenation of current dictionary matrices,
(ii)~null space identification, and (iii)~dictionary reduction, until
the desired dictionary is identified.
Algorithm~\ref{algo:ssd} presents the pseudocode\footnote{The function
  $\operatorname{null}([A_i,B_i])$ returns a basis for the null space
  of $[A_i,B_i]$, and $Z_i^A$ and $Z_i^B$ in Step~\ref{algossd:null}
  have the same size.}.
\begin{algorithm}
  \caption{Symmetric Subspace Decomposition} \label{algo:ssd}
  \begin{algorithmic}[1] 
    \State \textbf{Initialization} 
    \State $i \gets 1$, $A_1 \gets \dx$, $B_1 \gets \dy$, $\cssd \gets I_{N_d}$

    \While{1}
    \smallskip
    \State $\begin{bmatrix} Z_i^A \\ Z_i^B \end{bmatrix} \gets
    \operatorname{null}([A_i,B_i])$ \Comment{Basis for the null
      space}\smallskip \label{algossd:null}
    \If{$\operatorname{null}([A_i,B_i])=\emptyset$}
    \State \textbf{return} $0$ \Comment{The basis does not exist}
    \State \textbf{break}
    \EndIf
   
    \If{$\rown(Z_i^A)  \leq \coln(Z_i^A)$} \label{algossd:size-check}
    \State \textbf{return} $\cssd$ \Comment{The procedure is complete} \label{algossd:retun-complete}
    \State \textbf{break}
    \EndIf
    \State $\cssd \gets \cssd Z_i^A$ \Comment{Reducing the subspace}
    \State $A_{i+1} \gets A_i Z_i^A$, $B_{i+1} \gets B_i Z_i^A$, $i \gets i+1$

    \EndWhile
  \end{algorithmic}
\end{algorithm}

\subsection{Convergence Analysis of the SSD Algorithm}

Here we characterize the convergence properties of the SSD
algorithm. The next result characterizes the dimension, maximality,
and symmetry of the subspace defined by its output.

\begin{theorem}\longthmtitle{Properties of SSD
    Output}\label{t:ssd-convergence} 
  Suppose Assumption~\ref{a:full-rank} holds.  For matrices $\dx,\dy$,
  let $ \cssd = \ssd(\dx,\dy)$.  The SSD algorithm has the following
  properties:
  \begin{enumerate}
  \item it stops after at most $N_d$ iterations;
  \item the matrix $ \cssd$ is either $0$ or has full column
    rank, and satisfies $\range(\dx \cssd) = \range(\dy
    \cssd)$;
    
  \item the subspace $\range(\dx \cssd)$ is maximal, in the sense that,
    for any matrix $E$ with $\range(\dx E) = \range(\dy E)$, we have
    $\range(\dx E) \subseteq \range(\dx \cssd)$ and $\range(E) \subseteq
    \range(\cssd)$;
  \item $\range\big(\ssd(\dx,\dy)\big) = \range\big(\ssd(\dy,\dx)\big)$.
  \end{enumerate}
\end{theorem}
\begin{proof}
  (a) First, we use (strong) induction to prove that at each iteration
  $Z_i^A,Z_i^B$ are matrices with full column rank \emph{upon
    existence}. By Assumption~\ref{a:full-rank}, $A_1$ and $B_1$ have
  full column rank. Now, by using Lemma~\ref{l:subspace-intersection}
  one can derive that $Z_1^A$ and $Z_1^B$ have full column rank. Now,
  suppose that the matrices $Z_1^A,\ldots,Z_k^A$ and
  $Z_1^B,\ldots,Z_k^B$ have full column rank. Using
  Assumption~\ref{a:full-rank} one can deduce that $A_{k+1}=A_1 Z_1^A
  \dots Z_k^A$, $B_{k+1} =B_1 Z_1^A \dots Z_k^A$ have full column rank
  since they are product of matrices with full column rank. Using
  Lemma~\ref{l:subspace-intersection}, one can conclude that
  $Z_{k+1}^A$ and $Z_{k+1}^B$ have full column rank.
  
  Consequently, we have $\rown(Z_i^A) \geq \coln(Z_i^A)$. Hence,
  Step~\ref{algossd:size-check} of the SSD algorithm implies that the
  algorithm can only move to the next iteration if $\rown(Z_i^A) >
  \coln(Z_i^A)$, which means the number of columns in $A_{i+1}$ and
  $B_{i+1}$ decreases with respect to $A_i$ and~$B_i$.  Hence, the
  algorithm terminates after at most $N_d$ iterations since $A_1$ and
  $B_1$ have $N_d$ columns.
		
  (b) The $\cssd =0$ case is trivial. Suppose that the algorithm stops
  after $k$ iterations with nonzero $\cssd$. This means that $Z_k^A$
  and $Z_k^B$ are square full rank matrices. Also, by definition we
  have $A_k Z_k^A=-B_k Z_k^B$ which means that $A_k = -B_k Z_k^B
  (Z_k^A)^{-1}$. Noting that $Z_k^B (Z_k^A)^{-1}$ is a full rank
  square matrix, one can derive $\range(A_k)=\range(B_k)$. A closer
  look at the definitions shows that $A_k=\dx \cssd$ and $B_k = \dy
  \cssd$. Hence, $\range(\dx \cssd) = \range(\dy \cssd)$.  Moreover,
  $\cssd=Z_1^A \cdots Z_{k-1}^A$ and considering the fact that
  $Z_1^A,\ldots,Z_{k-1}^A$ have full column rank, one can deduce that
  $\cssd$ has full column rank.
  
  (c) Suppose that the matrix $E$ satisfies $\range(\dx E) =
  \range(\dy E)$. First, we use induction to prove that $\range(\dx E)
  \subseteq \range(A_i) \cap \range(B_i)$ for each iteration $i$ that
  the algorithm goes through. Let $i=1$, then $A_1=\dx$ and $B_1 =
  \dy$. Consequently, $\range(\dx E) \subseteq \range(A_1)$ and
  $\range(\dx E)=\range(\dy E) \subseteq \range(B_1)$ based on the
  definition of $E$. Hence, $\range(\dx E) \subseteq \range(A_1) \cap
  \range(B_1)$. Now, suppose
  \begin{align}\label{eq:induction-step-i}
    \range(\dx E) \subseteq \range(A_i) \cap \range(B_i).
  \end{align}
  Using Lemma~\ref{l:subspace-intersection}, one can derive
  $\range(A_i Z_i^A) = \range(A_i) \cap \range(B_i)$. Combining this
  with~\eqref{eq:induction-step-i}, we get
  \begin{align}\label{eq:subset1}
    \range(\dx E) \subseteq \range(A_i Z_i^A)
    = \range \big(\dx Z_1^A \cdots Z_i^A\big).
  \end{align}
  Using~\eqref{eq:subset1} with Lemma~\ref{l:product-subspace} one can
  derive $\range(E) \subseteq \range(Z_1^A \cdots Z_i^A)$. Using
  Lemma~\ref{l:product-subspace} once again, we get
  \begin{align}\label{eq:subset2}
    \range(\dx E) = \range(\dy E)
    \subseteq \range \big(\dy Z_1^A \cdots Z_i^A \big).
  \end{align}
  Definition of $A_{i+1},B_{i+1}$ along with~\eqref{eq:subset1}
  and~\eqref{eq:subset2} lead to conclusion that $\range(\dx E)
  \subseteq \range(A_{i+1}) \cap \range(B_{i+1})$ and the induction is
  complete.
		
  Now, suppose that the algorithm terminates at iteration $k$. In the
  case that $\cssd =0$, we have $\range(A_k) \cap \range(B_k) =\{0\}$,
  which means that $E=0$ and $\range(\dx E) \subseteq \range(\dx
  \cssd)$. In the case that $\cssd \neq 0$, using the fact that
  $\range(\dx E) \subseteq \range(A_k) \cap \range(B_k)$, $\cssd =
  Z_1^A \cdots Z_{k-1}^A$, and $\range(\dx \cssd) = \range(\dy
  \cssd)$, one can deduce that $\range(\dx E) \subseteq \range(\dx
  \cssd)$. Moreover, using Assumption~\ref{a:full-rank} and
  Lemma~\ref{l:product-subspace} one can write $\range(E) \subseteq
  \range(\cssd)$.
  
  (d) For convenience, let $E^\ssd = \ssd(D(Y),D(X))$. Based on the
  definition of $\cssd$ and $E^\ssd$, one can write
  \begin{align*}
    \range(D(X) \cssd) = \range(D(Y) \cssd
    ) 
    \\
    \range(D(X) E^\ssd) = \range(D(Y) E^\ssd
    ) 
  \end{align*}
  These equations in conjunction with the maximality of
  $\range(\cssd)$ from part~(c) imply $\range(E^\ssd) \subseteq
  \range(\cssd)$. Using a similar argument, invoking the maximality of
  $\range(E^\ssd)$, we have $\range(\cssd) \subseteq \range(E^\ssd)$,
  concluding the~proof.
\end{proof}

\begin{remark}\longthmtitle{Time and Space Complexity of the SSD
    Algorithm}\label{r:ssd-complexity}
  Given $N$ data snapshots and a dictionary with $N_d$ elements, where
  usually $N \gg N_d$, and assuming that operations on scalar elements
  require time and memory of order $O(1)$, the most time and memory
  consuming operation in the SSD algorithm is
  Step~\ref{algossd:null}. This step can be done by truncated Singular
  Value Decomposition~(SVD) and finding the perpendicular space to the
  span of the right singular vectors, with time complexity $O(NN_d^2)$
  and memory complexity $O(NN_d)$, see
  e.g.,~\cite{XL-SW-YC:19}. Since, based on
  Theorem~\ref{t:ssd-convergence}(a), the SSD algorithm terminates in
  at most $N_d$ iterations, the total time complexity is $O(N
  N_d^3)$. However, since at each iteration we can reuse the memory
  for Step~\ref{algossd:null}, the space complexity of SSD is $O(N
  N_d)$.  \oprocend
\end{remark}

Note that SSD removes the functions that do not evolve linearly in
time according to the available data snapshots. Therefore, as we
gather more data, the identified subspace either remains the same or
gets smaller, as stated next.

\begin{lemma}\longthmtitle{Monotonicity of SSD Output with Respect to
    Data Addition}\label{l:ssd-monotone}
  Let $D(X), D(Y)$ and $D(\hat{X}),D(\hat{Y})$ be two pairs of data
  snapshots such that
  \begin{align}\label{eq:data-monotone}
    \rows \big( [D(X), D(Y)] \big) \subseteq \rows
    \big([D(\hat{X}),D(\hat{Y})] \big),
  \end{align}
  and for which Assumption~\ref{a:full-rank} holds. Then
  \begin{align*}
    \range( \ssd([D(\hat{X}),D(\hat{Y})])) \subseteq \range(\ssd(D(X),
    D(Y))) .
  \end{align*}
\end{lemma}
\begin{proof}
  We use the shorthand notation $\hat{C} =
  \ssd([D(\hat{X}),D(\hat{Y})])$ and $C = \ssd(D(X), D(Y))$.
  From~\eqref{eq:data-monotone}, we deduce that there exists a matrix
  $E$ with $\rows(E) \subseteq \rows(I_{\rown(\hat{X})})$ such that
  \begin{align}\label{eq:data-inclusion}
    E D(\hat{X}) = D(X) , \; E D(\hat{Y}) = D(Y).
  \end{align}
  Moreover, based on the definition of $\hat{C}$ and
  Theorem~\ref{t:ssd-convergence}(b), we have $\range(D(\hat{X})
  \hat{C}) = \range(D(\hat{Y}) \hat{C})$. Hence, there exists a full
  rank square matrix $\hat{K}$ such that
  \begin{align*}
    D(\hat{Y}) \hat{C} = D(\hat{X}) \hat{C} \hat{K}.
  \end{align*}
  Multiplying both sides from the left by $E$ and
  using~\eqref{eq:data-inclusion} gives $D(Y) \hat{C} = D(X) \hat{C}
  \hat{K}$. Consequently, we have $\range(D(Y) \hat{C}) = \range(D(X)
  \hat{C})$.  Now, the maximality of $C$
  (Theorem~\ref{t:ssd-convergence}(c)) implies $\range(\hat{C})
  \subseteq \range(C)$.
\end{proof}

\begin{remark}\longthmtitle{Implementing SSD on Finite-Precision Machines}\label{r:finite-precision-approx} 
  Since SSD is iterative, its implementation using finite precision
  leads to small errors that can affect the rank and null space of
  $[A_i,B_i]$ in Step~\ref{algossd:null}. To circumvent this issue,
  one can approximate $[A_i,B_i]$ at each iteration by a close (in the
  Frobenius norm) low-rank matrix. Let $\sigma_1 \geq \ldots \geq
  \sigma_{l_i}$ be the singular values of $[A_i, B_i] \in \real^{N
    \times l_i}$. Given a design parameter $\epsilon >0$, let
    $k_i$ be the minimum integer such that
  \begin{align}\label{eq:low-rank-approx-singular-values}
    \sum_{j=k_i}^{l_i} \sigma_j^2 \leq \epsilon \big(\sum_{j=1}^{l_i} \sigma_j^2 \big).
  \end{align}
  One can then  construct the matrix $[\hat{A}_i,\hat{B}_i]$ by setting
  $\sigma_{k_i}=\cdots=\sigma_{l_i}=0 $ in the singular value
  decomposition of $[A_i,B_i]$. The resulting matrix has lower rank and
  \begin{align}\label{eq:low-rank-approx-Fro-error}
    \|[A_i,B_i] - [\hat{A}_i,\hat{B}_i] \|_F^2 \leq \epsilon \|[A_i,B_i]\|_F^2.
  \end{align}
  Hence, $\epsilon$ tunes the accuracy of the approximation.  It is
  important to note that similar error bounds can be found for other
  unitarily invariant norms, see e.g.~\cite{LM:60}.
  \oprocend
\end{remark}

\subsection{Identification of Linear Evolutions and Koopman
    Eigenfunctions with the SSD Algorithm}

  Here we study the properties of the output of the SSD algorithm in
  what concerns the identification of the maximal Koopman-invariant
  subspace and the Koopman eigenfunctions. If $\cssd \neq 0$, we
define the invariant dictionary as
\begin{align}\label{eq:new-dictioanry-definition}
  \tilde{D}(x) := D(x) \cssd.
\end{align}
\new{To find the action of the Koopman operator on the subspace
  spanned by $\tilde{D}$, we apply EDMD on $\tdx$ and $\tdy$ to find
  \begin{align}\label{eq:Kssd-closed-form}
    \Kssd &= \EDMD{\tilde{D}}{X}{Y} = \tilde{D}(X)^\dagger \tilde{D}(Y)
    \notag \\
    &= \big( D(X) \cssd \big)^\dagger \big(D(Y) \cssd).
  \end{align}
  Based on Theorem~\ref{t:ssd-convergence}(b), we have
  \begin{align}\label{eq:new-dictionary-range-equality}
    \range(\tilde{D}(X)) = \range(\tilde{D}(Y)).
  \end{align}
  Moreover, $\tilde{D}(X)$ and $\tilde{D}(Y)$ have full column rank as a
  result of Assumption~\ref{a:full-rank} and
  Theorem~\ref{t:ssd-convergence}(b). Consequently, $\Kssd$ is a (unique) nonsingular matrix satisfying
\begin{align}\label{eq:new-dictionary-Kssd}
\tilde{D}(Y) = \tilde{D}(X) \Kssd.
\end{align}
Interestingly, equation~\eqref{eq:new-dictionary-Kssd} implies that
the residual error of EDMD, $\| \tdy - \tdx \Kssd\|_F$, is equal to
zero. }
Based on~\eqref{eq:new-dictionary-range-equality}, one can find
$\Kssd$ \new{more efficiently and} only based on partial data instead
of calculating the pseudo-inverse of $D(X)\cssd$. Formally, consider
full column rank data matrices $D(\hat{X}), D(\hat{Y})$ such that
\begin{align*}
  \rows[D(\hat{X}), D(\hat{Y})] \subseteq \rows[D(X),D(Y)].
\end{align*}
Then, \new{$ \Kssd = \EDMD{\tilde{D}}{\hat{X}}{\hat{Y}}$}.
Next, we show that the eigenvectors of
$\Kssd$ fully characterize the functions that evolve linearly in time
according to the available data.

\begin{theorem}\longthmtitle{Identification of Linear Evolutions using
    the SSD Algorithm}\label{t:linear-evolutions-ssd-algorithm}
  Suppose that Assumption~\ref{a:full-rank} holds. Let $\cssd =
  \ssd(D(X),D(Y)) \neq 0$, $\Kssd =\big( D(X) \cssd \big)^\dagger
  \big(D(Y) \cssd \big)$, and $f(x) \in \Span(D(x))$ denoted as $f(x)
  = D(x)v$ with $v \in \cplx^{N_d}\setminus \{0\}$.
  Then $f(y_i) = \lambda f(x_i)$ for some $\lambda \in \cplx \setminus
  \{0\}$ and for all $i \in \until{\rown(X)}$ if and only if $v=\cssd
  w$ with $\Kssd w = \lambda w$.
\end{theorem}
\begin{proof}
  $(\Leftarrow)$: Based on definition of $\Kssd$,
  Assumption~\ref{a:full-rank}, and considering the fact that $\cssd$
  has full column rank (Theorem~\ref{t:ssd-convergence}(b)), one can
  use~\eqref{eq:new-dictioanry-definition}-\eqref{eq:new-dictionary-Kssd}
  and the fact that $\Kssd w = \lambda w$ to write $\dy \cssd w =
  \lambda \dx \cssd w$. Consequently, using $v = \cssd w$ we have
  \begin{align*}
    \dy v = \lambda \dx v.
  \end{align*}
  By inspecting the equation above in a row-wise manner, one can
  deduce that $f(y_i) = \lambda f(x_i)$ for some $\lambda \in \cplx
  \setminus \{0\}$ and for all $i \in \until{\rown(X)}$, as claimed.

  $(\Rightarrow)$: Based on the hypotheses, we have
  \begin{align}\label{eq:dictionary-linear-evo}
    D(Y)v = \lambda D(X)v.
  \end{align}
  Consider first the case when $v \in \real^{N_d}$. Then
  using~\eqref{eq:dictionary-linear-evo}, we deduce $\range(\dx v) =
  \range(\dy v)$. The maximality of $\cssd$
  (Theorem~\ref{t:ssd-convergence}(c)) implies that $\range(v)
  \subseteq \range(\cssd)$ and consequently $v=\cssd w$ for some $w$.
  Replacing $v$ by $\cssd w$ in~\eqref{eq:dictionary-linear-evo} and
  using the definition of $\Kssd$, one deduces $\Kssd w = \lambda w$.

  Now, suppose that $v = v_R + j v_I$ with $v_I \neq 0$. Since $\dx$
  and $\dy$ are real matrices, one can
  use~\eqref{eq:dictionary-linear-evo} and write $\dy \bar{v} =
  \bar{\lambda} \dx \bar{v}$. This, together
  with~\eqref{eq:dictionary-linear-evo}, implies
  \begin{align}\label{eq:complex-conjugate}
    \dy E = \dx E \Lambda,
  \end{align}
  where $E = [v_R, v_I]$ and
  \begin{align*}
    \Lambda = \begin{bmatrix} \re(\lambda) & \im(\lambda) \\
      -\im(\lambda) &\re(\lambda) \end{bmatrix}.
  \end{align*}
  Since $\Lambda$ is full rank, we have $\range(\dx E) = \range(\dy
  E)$ and using Theorem~\ref{t:ssd-convergence}(c), one can conclude
  $\range(E) \subseteq \range(\cssd)$. Consequently, there exists a
  real vector $z$ such that $E = \cssd z$. By replacing this
  in~\eqref{eq:complex-conjugate} and multiplying both sides from the
  right by $r=[1,j]^T$ and defining $w = z r$, one can conclude that
  $v=Er = \cssd w$ and $\dy \cssd w = \lambda \dx \cssd w$. This in
  conjunction with the definition of $\Kssd$ implies that $\Kssd w =
  \lambda w$,  concluding the proof.
\end{proof}

Using Theorem~\ref{t:linear-evolutions-ssd-algorithm}, one can
identify all the linear evolutions in the span of the original
dictionary, thereby establishing an equivalence with the
forward-backward EDMD characterization of
Section~\ref{sec:EDMD-charac}.

\begin{corollary}\longthmtitle{Equivalence of Forward-Backward EDMD
    and SSD in the Identification of Linear
    Evolutions}\label{c:f-b-EDMD=SSD} 
  Suppose that Assumption~\ref{a:full-rank} holds. Let $K_f =
  \EDMD{D}{X}{Y}$, $K_b= \EDMD{D}{Y}{X}$, $\cssd =
  \ssd(D(X),D(Y)) \neq 0$ and $\Kssd =\big( D(X) \cssd \big)^\dagger
  \big(D(Y) \cssd \big)$. Then, $K_f v = \lambda v$ and $K_b v =
  \lambda^{-1} v$ for some $v \in \cplx^{N_d}\setminus \{0\}$ and
  $\lambda \in \cplx \setminus \{0\}$ if and only if there exists
  vector $w$ such that $v = \cssd w$ and $\Kssd w = \lambda w$.
\end{corollary}

The proof of this result is a consequence of
Theorems~\ref{t:forward-backward-EDMD-evo}
and~\ref{t:linear-evolutions-ssd-algorithm}.  Note that the linear
evolutions identified by SSD might not be Koopman eigenfunctions,
since we can only guarantee that they evolve linearly according to the
available data snapshots, not starting everywhere in the state space
$\Mc$. The following result uses the equivalence between SSD and the
Forward-Backward EDMD method to provide a guarantee for the
identification of Koopman eigenfunctions.

\begin{theorem}\longthmtitle{Identification of Koopman Eigenfunctions
    by the SSD Algorithm}\label{t:eig-identification-ssd} 
  Given an infinite sampling, suppose that the sequence of dictionary
  snapshot matrices is $R$-rich. For $N \geq R$, let $\cssd_N =
  \ssd(D(X_{1:N}),D(Y_{1:N})) \neq 0$, and $\Kssd_N =\big( D(X_{1:N})
  \cssd_N \big)^\dagger \big(D(Y_{1:N}) \cssd_N)$. Given $v \in
  \cplx^{N_d}\setminus\{0\}$ and $\lambda \in \cplx \setminus\{0\}$,
  let $f(x)=D(x)v$. Then,
  \begin{enumerate}
  \item If $f$ is an eigenfunction of the Koopman operator with
    eigenvalue $\lambda$, then for every $N \geq R$, there exists
    $w_N$ such that $v =\cssd_N w_N$ and $K_N^\ssd w_N = \lambda w_N$;
  \item Conversely, and assuming the dictionary functions are
    continuous and Assumption~\ref{a:dense-sampling} holds, if $v \in
    \range(\cssd_N)$ and there exists $w_N$ such that $v =\cssd_N w_N$
    and $\Kssd_N w_N = \lambda w_N$ for every $N \geq R$, then $f$ is
    an eigenfunction of the Koopman operator with probability~1.
  \end{enumerate}
\end{theorem}

This result is a consequence of
Theorem~\ref{t:forward-backward-EDMD-eig} and
Corollary~\ref{c:f-b-EDMD=SSD}.
Theorem~\ref{t:eig-identification-ssd} shows that the SSD algorithm
finds all the eigenfunctions in the span of the original dictionary
almost surely. The identified eigenfunctions span a Koopman-invariant
subspace. This subspace however is not necessarily the maximal
Koopman-invariant subspace in the span of the original
dictionary. Next, we show that the SSD method actually identifies the
maximal Koopman-invariant subspace in the span of the dictionary.

\begin{theorem}\longthmtitle{SSD Finds the Maximal Koopman-Invariant
    Subspace as $N \to \infty$}\label{t:ssd-limit-maximal}
  Given an infinite sampling and a dictionary composed of continuous
  functions, suppose that the sequence of dictionary snapshot matrices
  is $R$-rich and Assumption~\ref{a:dense-sampling} holds.  Let the columns of 
  $\csinf$ form a basis for $\lim_{N \to \infty} \range(\cssd_{N})$, i.e.,
  \begin{align}\label{eq:c-inf-def}
    \range(\csinf) &= \lim_{N \to \infty} \range(\cssd_{N}) =
    \bigcap_{N=R}^\infty \range(\cssd_N).
  \end{align}
  (note that the sequence $\{\range(\cssd_{N})\}_{N=1}^{\infty}$ is
  monotonic, and hence convergent).
  Then $\Span(D(x)\csinf)$ is the maximal Koopman-invariant subspace
  in the span of the dictionary~$D$ with probability 1.
\end{theorem}
\begin{proof}
  If $\csinf = 0$, considering the fact that for all $N \geq R$,
  $\range(\cssd_{N+1}) \subseteq \range(\cssd_N)$
  (Lemma~\ref{l:ssd-monotone}), one deduces that there exists $m \in
  \naturals$ such that for all $i \ge m$, $\cssd_i = 0$. Hence based
  on Theorem~\ref{t:ssd-convergence}(c), the maximal Koopman-invariant
  subspace acquired from the data is $\{0\}$.  Noting that the
  subspace identified by SSD contains the maximal Koopman-invariant
  subspace, we deduce that the latter is the zero subspace, which is
  indeed spanned by $D(x)\csinf$.

  Now, suppose that $\csinf \neq 0$ and has full column rank.  First,
  we show that
  \begin{align}\label{eq:ssd-inv-1}
    \range(D(X_{1:N}) \csinf) = \range(D(Y_{1:N}) \csinf), \quad
    \forall N \geq R .
  \end{align}
  Considering~\eqref{eq:c-inf-def} and the fact that for all $N \geq
  R$, $\range(\cssd_{N+1}) \subseteq \range(\cssd_N)$, we can write
  for all $N \geq R$
  \begin{align*}
    \range(\csinf) = \bigcap_{i=N}^\infty \range(\cssd_i).
  \end{align*}
  Invoking Lemma~\ref{l:infinite-intersection-product}, we have for
  all $N \geq R$,
  \begin{subequations}\label{eq:ssd-inv-2}
    \begin{align}
      \range(D(X_{1:N}) \csinf) & = \bigcap_{i=N}^{\infty}
      \range(D(X_{1:N}) \cssd_i),
      \\
      \range(D(Y_{1:N}) \csinf) & = \bigcap_{i=N}^{\infty}
      \range(D(Y_{1:N}) \cssd_i).
    \end{align}
  \end{subequations}
  Moreover, for all $i \geq N$ we have $\range(D(X_{1:i})\cssd_i) =
  \range(D(Y_{1:i})\cssd_i)$ and hence by looking at this equality in
  a row-wise manner, one can write
  \begin{align}\label{eq:ssd-inv-3}
    \range(D(X_{1:N})\cssd_i) =\range(D(Y_{1:N})\cssd_i), \quad \forall i
    \geq N.
  \end{align}
  The combination of~\eqref{eq:ssd-inv-2} and~\eqref{eq:ssd-inv-3}
  yields~\eqref{eq:ssd-inv-1}. Based on the latter, the fact that
  $D(X_{1:N})$ and $D(Y_{1:N})$ have full column rank for every $N
  \geq R$ and the fact that $\csinf$ has full column rank, there
  exists a \emph{unique} nonsingular square matrix $\Ksinf \in
  \real^{\coln(\csinf) \times \coln(\csinf)}$
  such that
  \begin{align}\label{eq:ssd-inv-4}
    D(X_{1:N})\csinf \Ksinf = D(Y_{1:N}) \csinf, \quad \forall N \geq
    R.
  \end{align}
  Note that $\Ksinf$ does not depend on $N$.  Next, we aim to prove
  that for every function $f \in \Span(D(x) \csinf)$, $\Kc(f)$ is also
  in $\Span(D(x)\csinf)$ almost surely. Let $v \in
  \real^{\coln(\csinf)}$ such that $ f(x) = D(x) \csinf v $ and define
  \begin{align}\label{eq:ssd-inv-5}
    g(x) := D(x) \csinf \Ksinf v.
  \end{align}
  We show that $g = f \circ T= \Kc(f)$ almost surely. Define the
  function $h := g - f \circ T$. Also, let $S_\infty =
  \bigcup_{N=R}^\infty S_N$ be the set of initial conditions. Based
  on~\eqref{eq:ssd-inv-4},~\eqref{eq:ssd-inv-5}, and definition of
  $h$,
  \begin{align*}
    h(x) = 0, \quad \forall x \in S_\infty.
  \end{align*}
  Moreover, $h$ is continuous since $D$ and $T$ are continuous. This,
  together with the fact that $S_\infty$ is dense in $\Mc$ almost
  surely (Assumption~\ref{a:dense-sampling}), we deduce $h \equiv 0$
  on $\Mc$ almost surely. Therefore, $g = \Kc(f) = f \circ T$ with
  probability 1.  Noting that $g(x) \in \Span(D(x) \csinf)$, we have
  proven that $\Span(D(x) \csinf)$ is Koopman invariant almost surely.

  Finally, we prove the maximality of $\Span(D(x) \csinf)$. Let $\Lc$
  be a Koopman-invariant subspace in $\Span(D(x))$. Then there exists
  a full column rank matrix $E$ such that $\Lc =
  \Span(D(x)E)$. Moreover, since the invariance of $\Lc$ reflects in
  data, $ \range(D(X_{1:N}) E) = \range(D(Y_{1:N}) E)$, for all $N
  \geq R$.  As a result, based on Theorem~\ref{t:ssd-convergence}(c),
  we have $ \range(E) \subseteq \range(\cssd_N)$, for all $N \geq R$,
  and hence $\range(E) \subseteq \range(\csinf)$. Therefore, by
  Lemma~\ref{l:product-subspace}, we have $ \Lc = \range(D(x) E)
  \subseteq \range(D(x)\csinf)$, which completes the proof.
\end{proof}

\begin{remark}\longthmtitle{Generalized Koopman
    Eigenfunctions}\label{r:generalized-eigs}
  One can also extend the above discussion for generalized Koopman
  eigenfunctions (see e.g.~\cite[Remark 11]{MB-RM-IM:12}). Given a
  generalized eigenvector $w$ of $\Kssd$, the corresponding
  generalized Koopman eigenfunction is $\phi(x) = D(x) \cssd w$.
  \oprocend
\end{remark}

\section{Streaming 
  Symmetric Subspace Decomposition}\label{sec:SSSD}

In this section, we consider the setup where data becomes available in
a streaming fashion. A straightforward algorithmic solution for this
scenario would be to re-run, at each timestep, the SSD algorithm with
all the data available up to then. However, this approach does not
take advantage of the answers computed in previous timesteps, and
maybe become inefficient when the size of the data is large. Instead,
here we pursue the design of an online algorithm, termed Streaming
Symmetric Subspace Decomposition (SSSD),
cf. Algorithm~\ref{algo:sssd}, that updates the identified subspaces
using the previously computed ones.  Note that the SSSD algorithm is
not only useful for streaming data sets but also for the case of
non-streaming large data sets for which the execution of SSD requires
a significant amount of memory.

\begin{algorithm}
  \caption{Streaming Symmetric Subspace
    Decomposition} \label{algo:sssd}
  \begin{algorithmic}[1] 
    \State \textbf{Initialization} 
    \smallskip
    \State $\dsx(1) \gets \begin{bmatrix} D(X_{1:S}) \\ D(x_{S+1})\end{bmatrix}$, 
    $\dsy(1) \gets \begin{bmatrix} D(Y_{1:S}) \\ D(y_{S+1})\end{bmatrix}$
    \smallskip
    
    \State $i \gets 1$, $A_1 \gets \dsx(1)$, $B_1 \gets \dsy(1)$, $C_0 \gets I_{N_d}$
    \While{1}
    \If{$C_{i-1}=0$}\smallskip
    \State $C_{i} \gets 0$ \Comment{The basis does not exist} \label{algosssd:zero}
    \State \textbf{return} $C_i$
    \State \textbf{break}
    \EndIf
    \State $F_i \gets \ssd(A_i,B_i)$    \label{algosssd:ssd}
    \If{$F_i=0$}\smallskip
    \State $C_{i} \gets 0$ \Comment{The basis does not exist} \label{algosssd:zero2}
    \State \textbf{return} $C_i$
    \State \textbf{break}
    \EndIf
    \If{$\rown(F_i) > \coln(F_i)$} \label{algosssd:reductioncheck}
    \State  $C_i \gets \basis(\range(C_{i-1}F_i))$ \Comment{Subspace reduction} \label{algosssd:basis}
    \Else    
    \State $C_i \gets C_{i-1}$  		 \Comment{No change} \label{algosssd:nochange}
    \EndIf
    \State \textbf{return} $C_i$
    \State $i \gets i+1$
    
    \Statex  \hspace{14pt}$\triangledown$ Replacing the last data snapshot with the new one
    \smallskip
    \State 
    $\dsx(i) = \begin{bmatrix} D(X_{1:S}) \\ D(x_{S+i})\end{bmatrix}$, 
    $\dsy(i) = \begin{bmatrix} D(Y_{1:S}) \\ D(y_{S+i})\end{bmatrix}$
    \smallskip
    \Statex  \hspace{14pt}$\triangledown$ Calculating the reduced dictionary snapshots
    \State $A_{i} \gets \dsx(i)C_{i-1}$, $B_{i} \gets \dsy(i)C_{i-1}$
    
    \EndWhile
  \end{algorithmic}
\end{algorithm}

Given the \emph{signature} snapshot matrices $X_{1:S}$ and $Y_{1:S}$,
for some $S \in \naturals$, and a dictionary of functions $D$, the
SSSD algorithm proceeds as follows: at each iteration, the algorithm
receives a new pair of data snapshots, combines them with signature
data matrices, and applies the latest available dictionary on
them. Then, it uses SSD on those dictionary matrices and further
prunes the dictionary. The basic idea of the SSSD algorithm stems
  from the monotonicity of SSD's output dictionary versus the data
  (cf. Lemma~\ref{l:ssd-monotone}), i.e., by adding more data the
  dimension of the dictionary does not increase. Since the SSD
algorithm relies on Assumption~\ref{a:full-rank}, we make the
following assumption on the signature snapshots and the original
dictionary.

\begin{assumption}\longthmtitle{Full Rank Signature Dictionary
    Matrices} \label{a:signature}
  We assume that there exists $S \in \naturals$ such that the matrices
  $D(X_{1:S})$ and $D(Y_{1:S})$ have full column rank.  \oprocend
\end{assumption}

For a finite number of data snapshots, Assumption~\ref{a:signature} is
equivalent to Assumption~\ref{a:full-rank}. For an infinite sampling,
Assumption~\ref{a:signature} holds for a $R$-rich sequence of snapshot
matrices.  The next result discusses the basic properties of the SSSD
output at each iteration.

\begin{proposition}\longthmtitle{Properties of SSSD
    Output}\label{p:sssd-properties} 
  Suppose Assumption~\ref{a:signature} holds. For $i \in
  \naturals$, let $C_i$ denote the output of the SSSD algorithm
  at the $i$th iteration. Then, for all $i \in
    \naturals$,
  \begin{enumerate}
  \item $C_i$ has full column rank or is equal to zero;
  \item $\range(C_i) \subseteq \range(C_{i-1})$;
  \item $\range(\dsx(i)C_i) = \range(\dsy(i) C_i)$.
  \end{enumerate}
\end{proposition}
\begin{proof}
  (a) We prove the claim by induction. $C_0 = I_{N_d}$ and has full
  column rank. Now, suppose that $C_k$ has full column rank or is
  zero. We show the same fact for $C_{k+1}$. If $C_k=0$, then SSSD
  executes Step~\ref{algosssd:zero} and we have $C_{k+1}=0$. Now,
  suppose that $C_k$ has full column rank. Considering the fact that
  $\dsx(k+1)$ and $\dsy(k+1)$ have full column rank, one can deduce
  that $A_{k+1}$ and $B_{k+1}$ have full column rank. Consequently,
  based on Theorem~\ref{t:ssd-convergence}(b), $F_{k+1}$ has full
  column rank or is equal to zero. In the former case, the algorithm
  executes Step~\ref{algosssd:basis} or Step~\ref{algosssd:nochange},
  and based on definition of $\basis$ function and the fact that $C_k$
  has full column rank, one deduces that $C_{k+1}$ has full column
  rank. In the latter case, the algorithm executes
  Step~\ref{algosssd:zero2}, and $C_{k+1}=0$, as claimed.

  Now we prove (b). Note that at iteration $i$, $C_i$ will be
  determined by either
  Step~\ref{algosssd:zero},~\ref{algosssd:zero2},~\ref{algosssd:nochange},
  or~\ref{algosssd:basis}. The proof for the first three cases is
  trivial. We only need to prove the result for the case when the SSSD
  algorithm executes Step~\ref{algosssd:basis}. Based on
  Theorem~\ref{t:ssd-convergence}(b), one can deduce that $F_i$ has
  full column rank. Also, we have $\range(F_i) \subseteq
  \range(I_{\coln(C_{i-1})})$. Hence using Step~\ref{algosssd:basis}
  and Lemma~\ref{l:product-subspace}, one can write
  \begin{align*}
    \range(C_i) = \range(C_{i-1} F_i) \subseteq \range(C_{i-1}
    I_{\coln(C_{i-1})} ) = \range(C_{i-1}),
  \end{align*}
  as claimed.

  Next, we prove part~(c). If the SSSD algorithm executes
  Step~\ref{algosssd:zero} or Step~\ref{algosssd:zero2}, then the
  result follows directly. Now, suppose that the algorithm executes
  Step~\ref{algosssd:basis} or Step~\ref{algosssd:nochange}.  Note
  that if the algorithm executes one of these two steps, then $F_i
  \neq 0$, $C_{i-1} \neq 0$ and they have full column rank
  (Theorem~\ref{t:ssd-convergence}(b)). Hence, $\rown(F_i) \geq
  \coln(F_i)$. As a result, if the algorithm executes
  Step~\ref{algosssd:nochange}, we have $\rown(F_i) = \coln(F_i)$ and
  consequently $\range(F_i) = \range(I_{\coln(C_{i-1})})$. Therefore,
  \begin{align}\label{eq:sssd-prop-1}
    \range(C_i)=\range(C_{i-1})=\range(C_{i-1} I_{\coln(C_{i-1})} ) =
    \range(C_{i-1} F_i).
  \end{align}
  Moreover, if the SSSD algorithm executes Step~\ref{algosssd:basis},
  then using the definition of $\basis$ function, we have
  \begin{align}\label{eq:sssd-prop-2}
    \range(C_i)=\range(C_{i-1} F_i).
  \end{align}
  Also, based on definition of $F_i$ at Step~\ref{algosssd:ssd},
  Theorem~\ref{t:ssd-convergence}(b), and the fact that $A_i = \dsx(i)
  C_{i-1}$ and $B_i = \dsy(i) C_{i-1}$, 
  \begin{align*}
    \range(\dsx(i) C_{i-1} F_i) = \range(\dsy(i) C_{i-1} F_i).
  \end{align*}
  Using this together with~\eqref{eq:sssd-prop-1} upon execution of
  Step~\ref{algosssd:nochange} and~\eqref{eq:sssd-prop-2} upon
  execution of Step~\ref{algosssd:basis}, one deduces $\range(\dsx(i)
  C_i) = \range(\dsy(i) C_i)$, concluding the proof.
\end{proof}

Next, we show that the SSSD algorithm at each iteration identifies
exactly the same subspace as the SSD algorithm given all the data up
to that iteration.

\begin{theorem}\longthmtitle{Equivalence of SSD and
    SSSD}\label{t:ssd-sssd-equivalence}
  Suppose Assumption~\ref{a:signature} holds.  For $i \in \naturals$,
  let $C_i$ denote the output of the SSSD algorithm at the $i$th
  iteration and let $\cssd_i = \ssd \big(D(X_{1:S+i}), D(Y_{1:S+i})
  \big)$. Then,
  \begin{align*}
    \range(C_i) = \range(\cssd_i), \quad \forall i \in \naturals .
  \end{align*}
\end{theorem}
\begin{proof}
  \emph{Inclusion $\range(\cssd_i) \subseteq \range(C_i)$ for all $i
    \in \naturals$:} We reason by induction.  Note that in the SSSD
  algorithm, for $i=1$ we have $F_1 = \cssd_1$. As a result, if $F_1 =
  0$ then based on Step~\ref{algosssd:zero2}, $C_1 = \cssd_1 = 0$. If
  the SSSD algorithm executes Step~\ref{algosssd:basis}, then using
  the fact that $C_0 = I_{N_d}$, one can write $\range(C_1) =
  \range(\cssd_1)$. Moreover, if the SSSD algorithm executes
  Step~\ref{algosssd:nochange}, based on
  Step~\ref{algosssd:reductioncheck} and
  Theorem~\ref{t:ssd-convergence}(b), one can deduce that
  $\range(\cssd_1) = \range(C_1) = \range(F_1) =
  \range(I_{N_d})$. Consequently, in all cases
  \begin{align}\label{eq:ssd-sssd-1-eq}
    \range(\cssd_1) = \range(C_1).
  \end{align}
  Hence, $\range(\cssd_1) \subseteq \range(C_1)$. Now, suppose that
  \begin{align}\label{eq:ssd-sssd-induction-1}
    \range(\cssd_k) \subseteq \range(C_k).
  \end{align}
  We need to show that $\range(\cssd_{k+1}) \subseteq
  \range(C_{k+1})$. If $\cssd_{k+1}=0$ then the proof follows. Now
  assume that $\cssd_{k+1} \neq 0$ and has full column rank based on
  Theorem~\ref{t:ssd-convergence}(b). By Lemma~\ref{l:ssd-monotone},
  we have
  \begin{align}\label{eq:ssd-sssd-induction-2}
    \range(\cssd_{k+1}) \subseteq \range(\cssd_k).
  \end{align}
  Using~\eqref{eq:ssd-sssd-induction-1}
  and~\eqref{eq:ssd-sssd-induction-2}, one can deduce
  $\range(\cssd_{k+1}) \subseteq \range(C_k)$. Consequently, based on
  the fact that $\cssd_{k+1} \neq 0$, we have $C_k \neq 0$ and hence
  has full column rank based on
  Proposition~\ref{p:sssd-properties}(a). Moreover, there exists a
  full column-rank matrix $E_k$ such that
  \begin{align}\label{eq:ssd-sssd-induction-3}
    \cssd_{k+1} = C_k E_k.
  \end{align}
  Two cases are possible. In case~1, the SSSD algorithm executes
  Step~\ref{algosssd:nochange}. In case~2, the algorithm executes
  Step~\ref{algosssd:zero2} or Step~\ref{algosssd:basis}.
  For case~1, we have $C_{k+1}=C_k$. Consequently,
  using~\eqref{eq:ssd-sssd-induction-3} and considering the fact that
  $\range(E_k) \subseteq \range(I_{\coln(C_k)})$ and the fact that
  $C_k$ has full column rank, one can use
  Lemma~\ref{l:product-subspace} and conclude
  \begin{align}\label{eq:ssd-sssd-induction-4}
    \range(\cssd_{k+1}) = \range(C_k E_k) \subseteq \range(C_k) = \range (C_{k+1}).
  \end{align}
  Now, consider case~2. In this case, we have
  \begin{align}\label{eq:ssd-sssd-induction-5}
    \range(C_{k+1}) = \range(C_k F_{k+1}).
  \end{align}
  Also, based on definition of $\cssd_{k+1}$ and
  Theorem~\ref{t:ssd-convergence}(b), one can write
  \begin{align*}
    \range(D(X_{1:k+1} \cssd_{k+1})) = \range(D(Y_{1:k+1}
    \cssd_{k+1})).
  \end{align*}
  Looking at this equation in a row-wise manner and considering the
  fact that $\rows \big( [\dsx(k+1), \dsy(k+1)] \big) \subseteq \rows
  \big( [D(X_{1:k+1}), D(Y_{1:k+1})] \big)$, one can write
  \begin{align*}
    \range(\dsx(k+1) \cssd_{k+1}) = \range(\dsy(k+1) \cssd_{k+1}).
  \end{align*}
  Now, using~\eqref{eq:ssd-sssd-induction-3} we have $\range(\dsx(k+1)
  C_k E_k) = \range(\dsy(k+1) C_k E_k)$. Also, noting the definition
  of $F_{k+1}$ and the fact that $A_k = \dsx(k+1) C_k$, $B_k =
  \dsy(k+1) C_k$, one can use Theorem~\ref{t:ssd-convergence}(c) to
  write $\range(E_k) \subseteq \range(F_{k+1})$. Since $C_k$ has
  full column rank, we
  use~\eqref{eq:ssd-sssd-induction-3},~\eqref{eq:ssd-sssd-induction-5},
  and Lemma~\ref{l:product-subspace} to write
  \begin{align}\label{eq:ssd-sssd-induction-6}
    \range(\cssd_{k+1}) = \range (C_k E_k) \subseteq \range (C_k
    F_{k+1}) = \range(C_{k+1}).
  \end{align}
  In both cases, equations~\eqref{eq:ssd-sssd-induction-4}
  and~\eqref{eq:ssd-sssd-induction-6} conclude the induction.

  \emph{Inclusion $\range(C_i) \subseteq \range(\cssd_i)$ for all $i
    \in \naturals$:} We reason by induction
  too. Using~\eqref{eq:ssd-sssd-1-eq}, we have $\range(C_1) \subseteq
  \range(\cssd_1)$. Now, suppose that
  \begin{align}\label{eq:ssd-sssd-induction-7}
    \range(C_k) \subseteq \range(\cssd_k).
  \end{align}
  We prove the same result for $k+1$. If $C_{k+1} = 0$ then the result
  directly follows. Now, assume that $C_{k+1} \neq 0$. Consequently,
  based on~\eqref{eq:ssd-sssd-induction-7},
  Proposition~\ref{p:sssd-properties}(a), and
  Theorem~\ref{t:ssd-convergence}(b), we deduce that $C_{k+1}$ and
  $\cssd_{k+1}$ have full column rank.
	
  The first part of the proof and~\eqref{eq:ssd-sssd-induction-7}
  imply that $\range(C_k) = \range(\cssd_k)$. Consequently, noting the
  fact that $\cssd_k$ is the output of the SSD algorithm with
  $D(X_{1:S+k})$ and $D(Y_{1:S+k})$, one can use
  Theorem~\ref{t:ssd-convergence}(b) to write
  \begin{align}\label{eq:ssd-sssd-induction-8}
    \range \big( D(X_{1:S+k})C_k \big) = \range \big( D(Y_{1:S+k})C_k
    \big).
  \end{align}
  Moreover, based on Proposition~\ref{p:sssd-properties}(b), we have
  $\range(C_{k+1}) \subseteq \range(C_k)$. Hence, since $C_k$ and
  $C_{k+1}$ have full column rank, there exists a matrix $G_k$ with
  full column rank such that
  \begin{align}\label{eq:ssd-sssd-induction-9}
    C_{k+1} = C_k G_k.
  \end{align}
  Also, based on Proposition~\ref{p:sssd-properties}(c) at iteration
  $k+1$ of the SSSD algorithm
  \begin{align}\label{eq:ssd-sssd-induction-rangek+1}
    \range(\dsx(k+1) C_{k+1}) = \range(\dsy(k+1) C_{k+1}).
  \end{align}
  Consequently, based on~\eqref{eq:ssd-sssd-induction-9}
  and~\eqref{eq:ssd-sssd-induction-rangek+1}, we have
  \begin{align*}
    \range(D(X_{1:S}) C_k G_k) = \range(D(Y_{1:S}) C_k G_k).
  \end{align*}
  Using this equation together with~\eqref{eq:ssd-sssd-induction-8} and
  Lemma~\ref{l:symmetric-immersion},
  \begin{align*}
    \range \big( D(X_{1:S+k})C_k G_k \big) = \range \big( D(Y_{1:S+k})C_k G_k\big).
  \end{align*}
  Moreover, using~\eqref{eq:ssd-sssd-induction-9} one can write
  \begin{align*}
    \range \big( D(X_{1:S+k})C_{k+1} \big) = \range \big( D(Y_{1:S+k})
    C_{k+1}\big).
  \end{align*}
  Hence, there exists a nonsingular square matrix $K^*$ such that
  \begin{align}\label{eq:ssd-sssd-induction-10}
    D(X_{1:S+k})C_{k+1} K^* = D(Y_{1:S+k}) C_{k+1}.
  \end{align}

  Also, based on~\eqref{eq:ssd-sssd-induction-rangek+1} and noting
  that $\dsx(k+1)$, $\dsx(k+1)$, and $C_{k+1}$ have full column rank,
  there exists a nonsingular square matrix $K$ such that
  \begin{align}\label{eq:ssd-sssd-induction-11}
    \dsx(k+1) C_{k+1} K = \dsy(k+1) C_{k+1}.
  \end{align}
  Using the first $S$ rows of~\eqref{eq:ssd-sssd-induction-10}
  and~\eqref{eq:ssd-sssd-induction-11}, one can write
  \begin{align*}
    D(X_{1:S}) C_{k+1} K^* = D(Y_{1:S}) C_{k+1},
    \\
    D(X_{1:S}) C_{k+1} K = D(Y_{1:S}) C_{k+1}.
  \end{align*}
  By subtracting the second equation from the first one, we get
  \begin{align*}
    D(X_{1:S}) C_{k+1} (K^*-K) = 0.
  \end{align*}
  Moreover, since $D(X_{1:S}) C_{k+1}$ has full column rank, we deduce
  $ K^* = K$.  Using this together
  with~\eqref{eq:ssd-sssd-induction-10}
  and~\eqref{eq:ssd-sssd-induction-11} yields
  \begin{align}\label{eq:ssd-sssd-induction-12}
    \range \big( D(X_{1:S+k+1})C_{k+1} \big) = \range \big( D(Y_{1:S+k+1}) C_{k+1}\big).
  \end{align}
  From~\eqref{eq:ssd-sssd-induction-12}, the definition of
  $\cssd_{k+1}$ and Theorem~\ref{t:ssd-convergence}(c), we deduce
  $\range(C_{k+1}) \subseteq \range(\cssd_{k+1})$, concluding the
  proof.
\end{proof}

Theorem~\ref{t:ssd-sssd-equivalence} establishes the equivalence
between the SSSD and SSD algorithms. As a consequence, all results
regarding the identification of Koopman-invariant subspaces and
eigenfunctions presented in Section~\ref{sec:SSD} are also valid for the
output of the SSSD algorithm.

\begin{remark}\longthmtitle{Time and Space Complexity of the
    SSSD Algorithm}\label{r:sssd-complexity}
  Given the first $N$ data snapshots and a dictionary with $N_d$
  elements, with $N > S \geq N_d$, and assuming that operations on
  scalar elements require time and space of order $O(1)$, the most
  time and memory consuming operation in the SSSD algorithm is
  Step~\ref{algosssd:ssd} invoking SSD. In this step, the most time
  consuming operation is performing SVD, with time complexity $O(S
  N_d^2)$ and space complexity of $O(S N_d)$, see
  e.g.,~\cite{XL-SW-YC:19}.  After having performed the first SVD, the
  ensuing ones result in a reduction of the dimension of the
  subspace. Therefore, the SSSD algorithm performs at most $N-S$ SVDs
  with no subspace reduction with overall time complexity $O(N S
  N_d^2)$ and at most $N_d$ SVD operations with subspace reductions
  with overall time complexity $O(S N_d^3)$. Considering the fact that
  $N \geq N_d$, the complexity of the SSSD algorithm is $O(N S
  N_d^2)$. Moreover, in many real world applications $S=O(N_d)$ (in
  fact usually $S=N_d$), which reduces the time complexity of SSSD to
  $O(N N_d^3)$, which is the same complexity as SSD.  Moreover, since
  we can reuse the space used in Step~\ref{algosssd:ssd} at each
  iteration, and considering the fact that the space complexity of
  this step is $O(SN_d)$, we deduce that the space complexity of SSSD
  is $O(S N_d)$. This usually reduces to $O(N_d^2)$ since $S=O(N_d)$
  in many real-world applications.
  \oprocend
\end{remark}

\begin{remark}\longthmtitle{SSSD is More Stable and Runs Faster than SSD}\label{r:sssd-faster-ssd}
  The SSSD algorithm is more computationally stable than SSD, since it
  always works with matrices of size at most $(S+1) \times N_d$ while
  SSD works with matrices of size $N \times N_d$. Moreover, even
  though SSD and SSSD have the same time complexity, the SSSD
  algorithm run faster for two reasons. First, at each iteration of
  the SSSD algorithm, the dictionary gets smaller, which reduces the
  cost of computation for the remaining data snapshots. Second, the
  characterizations in Remarks~\ref{r:ssd-complexity}
  and~\ref{r:sssd-complexity} only consider the number of floating
  point operations for the time complexity and ignore the amount of
  time used for loading the data. SSSD needs to load significantly
  smaller data matrices, which leads to a considerable reduction in
  run time compared to SSD.  \oprocend
\end{remark}

\section{Approximating Koopman-Invariant Subspaces}

We note that, if the span of the original dictionary~$D$ does not
contain any Koopman-invariant subspace, then the SSD algorithm returns
the trivial solution, which does not result in any information about
the behavior of the dynamical system.  To circumvent this issue, here
we propose a method to \emph{approximate} Koopman-invariant
subspaces. Noting the fact that the existence of a Koopman-invariant
subspace translates into the rank deficiency of the concatenated
matrix $[A_i, B_i]$ in Step~\ref{algossd:null} of the SSD algorithm,
we propose to replace the $\operatorname{null}$ function in SSD with
the $\operatorname{approx-null}$ routine presented in
Algorithm~\ref{algo:approx-null} below. This routine constructs an
approximated null space by selecting a set of small singular values.
The parameter $\epsilon > 0$ in Algorithm~\ref{algo:approx-null} is a
design choice that tunes the accuracy of the approximation\footnote{In
  Algorithm~\ref{algo:approx-null}, $A$ and $B$ have equal size and
  both have full column rank.}.

\begin{algorithm}
\caption{ $\operatorname{approx-null(A,B,\epsilon)}$} \label{algo:approx-null}
\begin{algorithmic}[1] 
{
	
\Statex  \hspace{0pt}$\triangledown$ Singular value decomposition of $[A,B]$
\State $\{U,S,V\} \gets \operatorname{svd}([A,B])$ \Comment $U S V^T = [A,B]$	
\smallskip

\State $ m \gets \coln(V)$ \Comment \# of columns of $V$ \label{algoan:size-V}
\vspace*{8pt}
\State $k_{\min} \gets \Big\{ \min_k \; 
\operatorname{s.t.}  \Big( \frac{\sum_{i=k}^m S_{i,i}^2}{ \| S \|_F^2} \leq \epsilon^ 2 \land  k > \coln(A) \Big) \Big\} $ \label{algoan:kmin}
\vspace*{8pt}

\Statex  \hspace{0pt}$\triangledown$ Choosing the right singular vectors corresponding to small singular values as the approximated null space
\If {$k_{\min} = \emptyset$} 
\State return $\emptyset$ \label{algoan:return-empty-no-k}
\State break  \label{algoan:break-empty-no-k}
\Else
\State $Z \gets (V_{k_{\min}:m}^T)^T$ \label{algoan:null-candidate}
\EndIf
\smallskip

\Statex \hspace{0pt}$\triangledown$ Make sure Assumption~\ref{a:full-rank} holds for the output
\While{1} \label{algoan:full-rank-procedure}
\smallskip
\State $\begin{bmatrix} Z^A \\ Z^B \end{bmatrix} \gets Z$ \Comment $\rown(Z^A) = \rown(Z^B)$ \label{algoan:Z-partition}
\smallskip

\If {$\rank{Z^A} = \rank{Z^B} = \coln(Z)$} \label{algoan:rank-check}
\State return $Z$ \Comment Basis for approximated null space \label{algoan:return-subspace}
\State break \label{algoan:break-complete}
\EndIf

\Statex  \hspace{12pt}$\triangledown$ Reducing the space
\If {$\coln(Z)=1$} \label{algoan:if-rank-one}
\State return $\emptyset$ \label{algoan:return-empty}
\State break \label{algoan:break-empty}
\Else 
\State $Z \gets (Z^T_{2:\coln(Z)})^T$ \Comment Removing the 1st column \label{algoan:size-reduction}
\EndIf

\EndWhile \label{algoan:full-rank-procedure-complete}
}
\end{algorithmic}
\end{algorithm}

The next result studies the basic properties of
Algorithm~\ref{algo:approx-null}.

\begin{proposition}\longthmtitle{Properties of
    Algorithm~\ref{algo:approx-null}}\label{p:approx-null-properties}
  Let $A$ and $B$ be matrices of equal size, $\epsilon >0$, and $Z =
  \operatorname{approx-null}(A,B,\epsilon)$. Then,
  \begin{enumerate}
  \item Algorithm~\ref{algo:approx-null} terminates in finite
    iterations;
  \item $Z$ is either $\emptyset$ or has full column rank;
  \item if $Z \neq \emptyset$, let $Z =[(Z^A)^T, (Z^B)^T]^T$ with
    $Z^A,Z^B$ of equal size.  Then $Z^A$ and $Z^B$ have full column
    rank.
  \end{enumerate}
\end{proposition}
\begin{proof}
  (a) We prove it by contradiction, i.e., suppose the algorithm does
  not terminate in finite iterations. Let $Z_i$ be the internal matrix
  in Step~\ref{algoan:null-candidate} at iteration $i$. Since by
  construction $k_{\min} > \coln(A)$ and $m = \coln(V) = \coln(A)+
  \coln(B)$ (cf. Step~\ref{algoan:size-V}), we deduce
  \begin{align}\label{eq:Z1-size}
    \coln(Z_1) = m - k_{\min} \leq \coln(B).
  \end{align}
  Moreover, since we assumed the algorithm never terminates, it
  executes Step~\ref{algoan:size-reduction} at each iteration and
  consequently, $\coln(Z_{i+1}) = \coln(Z_i)-1$ for $i \in
  \naturals$. As a result, one can use~\eqref{eq:Z1-size} to write
  \begin{align*}
    \coln(Z_j) = \coln(Z_1) - j+1 \leq \coln(B) -j + 1,
  \end{align*}
  which leads to $\coln(Z_j) < 0$ for $j > \coln(B) +1$, contradicting
  $\coln(Z_j) \geq 0$.

  (b) There are three ways for Algorithm~\ref{algo:approx-null} to
  terminate: either
  Steps~\ref{algoan:return-subspace}-\ref{algoan:break-complete},
  Steps~\ref{algoan:return-empty-no-k}-\ref{algoan:break-empty-no-k},
  or Steps~\ref{algoan:return-empty}-\ref{algoan:break-empty}. The
  latter two cases imply $Z = \emptyset$. In the other case, since the
  columns of $Z$ are selected from the right singular vectors of
  $[A,B]$, they are nonzero and mutually orthogonal. Consequently, $Z$
  has full column rank.

  (c) Since $Z \neq \emptyset$, the algorithm executes
  Steps~\ref{algoan:return-subspace}-\ref{algoan:break-complete} upon
  termination. Hence, the condition in Step~\ref{algoan:rank-check}
  holds, and consequently $Z^A$ and $Z^B$ have full column rank.
\end{proof}

We next characterize the quality of Algorithm~\ref{algo:approx-null}'s
output.

\begin{proposition}\longthmtitle{Quality of Low-Rank Approximation of
    $[A,B]$ Constructed with Output of
    Algorithm~\ref{algo:approx-null}}\label{p:approx-null-low-rank}
  Let $\epsilon > 0$, $A$ and $B$ full column rank matrices with equal
  size, and assume $Z = \operatorname{null-approx}(A,B,\epsilon) \neq
  \emptyset$. Denote $W = [A,B]$ and let $W = U S V^T$ be its singular
  value decomposition.  Let $\bar{S}$ be defined by setting in $S$
  the entries $\bar{S}_{i,i} = 0$ for $i \in \{\coln(V)-\coln(Z)+1,
  \ldots , \coln(V)\}$. Define $\bar{W} = U \bar{S} V^T$ and express
  it as the concatenation $\bar{W} = [\bar{A},\bar{B}]$, where
  $\bar{A}$ and $\bar{B}$ have the same size. Then,
  \begin{enumerate}
  \item $\|W-\bar{W}\|_F \leq \epsilon \|W\|_F$;
  \item the columns of $Z$ form a basis for the null space of
    $\bar{W}$;
  \item $\bar{A} Z^A = -\bar{B} Z^B$, where $Z = [(Z^A)^T, (Z^B)^T]^T$
    and $Z^A, Z^B$ have the same size.
  \end{enumerate}
\end{proposition}
\begin{proof}
  (a) By construction we have $ Z = V^T_{(\coln(V)-\coln(Z)+1 ) :
    \coln(V)}$, i.e., the columns of $Z$ are the last $\coln(Z)$
  columns of $V$, corresponding to the smallest singular values
  of~$W$. Moreover, based on Step~\ref{algoan:kmin} of the algorithm,
  the fact that the singular values are ordered in a decreasing manner
  in $S$, and noting that $k_{\min} \leq \coln(V) - \coln(Z)+1$, one
  can write
  \begin{align*}
    \sum_{i=\coln(V) - \coln(Z)+1}^{\coln(V)} S_{i,i}^2 \leq
    \epsilon^2 \sum_{i=1}^{\coln(V)} S_{i,i}^2 = \epsilon^2 \|W\|_F^2.
  \end{align*}
  The proof concludes by noting that the left hand side term in the
  previous equation is equal to $\|W - \bar{W}\|_F^2$.

  (b) The proof directly follows from the fact that $\bar{W} = U
  \bar{S} V^T$ is the singular value decomposition of $\bar{W}$ and
  the columns of $Z$ are the right singular vectors corresponding to
  zero singular values of $\bar{W}$.

  (c) Based on~(b), $\bar{W} Z = \mathbf{0}$. Hence, $\bar{A}
  Z^A = -\bar{B} Z^B$.
\end{proof}

We formally define the \emph{Approximated-SSD} algorithm as the
modification of SSD that replaces Step~\ref{algossd:null} of
Algorithm~\ref{algo:ssd} by
\begin{align*}
  \begin{bmatrix} Z_i^A \\ Z_i^B \end{bmatrix} \gets
  \operatorname{approx-null}(A_i,B_i,\epsilon).
\end{align*}
Since all other steps of Approximated-SSD are identical to SSD, we
omit presenting it for space reasons.

For convenience, we denote the output of the Approximated SSD
algorithm by $\cssdapprox$.
Proposition~\ref{p:approx-null-properties} completely preserves the
logical structure for the proof of Theorem~\ref{t:ssd-convergence}(a)
and, as a result, we deduce that the Approximated-SSD algorithm
terminates in at most $N_d$ iterations.  Moreover, the $\cssdapprox$
matrix is zero or has full column rank, since the second part of the
proof for Theorem~\ref{t:ssd-convergence}(b) also holds for
Approximated-SSD.  If $\cssdapprox \neq 0$, one can define the reduced
dictionary with $\tilde{N}_d = \coln(\cssdapprox)$ elements as
\begin{align}\label{eq:new-dictionary-approx}
  \Dtilde(x) = D(x) \cssdapprox , \; \forall x \in \Mc.
\end{align}
We propose calculating the linear prediction matrix $\Kssdapprox$ by
solving the following total least squares (TLS) problem (see
e.g.~\cite{IM-SVH:07} for more information on TLS)
\begin{subequations}\label{eq:TLS}
  \begin{align}
    & \underset{K,\Delta_1,\Delta_2}{\text{minimize}} &&
    \left\|[\Delta_1,\Delta_2]\right\|_F \label{eq:TLS-obj}
    \\
    & \text{subject to} && \Dtilde(Y)+\Delta_2= (\Dtilde(X)+\Delta_1)
    K. \label{eq:TLS-const}
  \end{align}
\end{subequations}
Even though TLS problems do not always have a solution, the next
result shows that~\eqref{eq:TLS} does. We also provide its closed-form
solution and a bound on the accuracy of the prediction \new{on the
  available data} based on the parameter~$\epsilon$.

\begin{theorem}\longthmtitle{Solution and Prediction Accuracy
    of~\eqref{eq:TLS}}\label{t:approximated-ssd-accuracy}
  Let $[\Dtilde(X),\Dtilde(Y)] = U S V^T$ be the singular value
  decomposition of $[\Dtilde(X),\Dtilde(Y)]$.  Let $\bar{S}$ be
  defined by setting in $S$ the entries $\bar{S}_{i,i} = 0$ for $i \in
  \{\tilde{N}_d +1, \ldots, 2\tilde{N}_d\}$. Let $ U \bar{S} V^T =
  [\bar{A}, \bar{B}] $, with $\bar{A}$, $\bar{B}$ of the same size.
  Define 
  \begin{subequations}
    \begin{align}
      \Kssdapprox & = \bar{A}^\dagger \bar{B}, \label{eq:Kapprox-def}
      \\
      [\Delta_1^*, \Delta_2^*] & = [\bar{A}, \bar{B}] - [\Dtilde(X),
      \Dtilde(Y)]. \label{eq:delta-star}
    \end{align}
  \end{subequations}
  Then, $\Kssdapprox$, $\Delta_1^*$, $\Delta_2^*$ are the global
  solution of~\eqref{eq:TLS} and
  \begin{align}
    \| [\Delta_1^*, \Delta_2^*] \|_F & \leq \epsilon \| [\Dtilde(X), \Dtilde(Y)]
    \|_F. \label{eq:prediction-accuracy}
  \end{align}
\end{theorem}
\begin{proof}
  One can rewrite~\eqref{eq:TLS-const} as 
  \begin{align*}
    \big( [\Dtilde(X), \Dtilde(Y)] + [\Delta_1, \Delta_2]
    \big) \begin{bmatrix} K
      \\
      -I_{\tilde{N}_d} \end{bmatrix} = 0,%
  \end{align*}
  which implies that $\rank{[\Dtilde(X), \Dtilde(Y)] + [\Delta_1,
    \Delta_2]} \leq \tilde{N}_d$. Using Eckart-Young
  theorem~\cite{CE-GY:36}, one deduces that $[\bar{A}, \bar{B}]$ is
  the closest matrix (in Frobenius norm) to $[\Dtilde(X), \Dtilde(Y)]$
  of rank smaller than or equal to $\tilde{N}_d$. In other words,
  $\Delta_1^*$ and $\Delta_2^*$ in~\eqref{eq:delta-star} minimize the
  cost function in~\eqref{eq:TLS-obj}. Next, we need to show that they
  also satisfy~\eqref{eq:TLS-const} with $\Kssdapprox$
  defined in~\eqref{eq:Kapprox-def}.

  Let $t$ be the termination iteration of the Approximated-SSD
  algorithm. Since $\cssdapprox \neq 0$, the algorithm executes
  Step~\ref{algossd:retun-complete}. Therefore, the condition in
  Step~\ref{algossd:size-check} holds and $\rown(Z^A_t) \leq
  \coln(Z^A_t)$, where $[(Z^A_t)^T, (Z^B_t)^T] =
  \operatorname{approx-null}(A_t,B_t, \epsilon)$. In addition, based
  on Proposition~\ref{p:approx-null-properties}(c), $Z^A_t$ and
  $Z^B_t$ are nonsingular square matrices. Noting that by definition
  in the Approximated-SSD algorithm, $A_t = \dx \cssdapprox =
  \Dtilde(X)$ and $B_t = \dy \cssdapprox = \Dtilde(Y)$, one can use
  Proposition~\ref{p:approx-null-low-rank}(c) with $W = [\Dtilde(X),
  \Dtilde(Y)]$ and $\bar{W} = [\bar{A},\bar{B}]$ to write $\bar{A}
  Z^A_t = - \bar{B} Z^B_t$. Since $Z^A_t$ and $Z^B_t$ are nonsingular
  square matrices, the previous equation leads to $\range(\bar{A}) =
  \range(\bar{B})$ and $\bar{A}\Kssdapprox = \bar{B}$, where
  $\Kssdapprox$ is defined in~\eqref{eq:Kapprox-def}. As a result,
  $\Delta_1^*, \Delta_2^*, \Kssdapprox$ satisfy the
  constraint~\eqref{eq:TLS-const}.  Finally, the accuracy bound
  defined in~\eqref{eq:prediction-accuracy} follows from
  Proposition~\ref{p:approx-null-low-rank}(a) with $W = [\Dtilde(X),
  \Dtilde(Y)]$ and $\bar{W} = [\bar{A}, \bar{B}]$.
\end{proof}

\new{Note that, unlike in the exact case
  (cf. Theorem~\ref{t:ssd-limit-maximal}),
  Theorem~\ref{t:approximated-ssd-accuracy} does not provide an
  out-of-sample bound on prediction accuracy.}  According to this
result, a small perturbation $ [\Delta_1^*, \Delta_2^*] $ to the
matrix $[\Dtilde(X), \Dtilde(Y)]$ allows us to describe the evolution
of the dictionary matrices linearly through~$\Kssdapprox$. Moreover,
the Frobenius norm of the perturbation is upper bounded by $\epsilon
\| [\Dtilde(X), \Dtilde(Y)] \|_F$, which implies that a smaller
$\epsilon$ leads to better accuracy \new{on the observed samples}.

\section{Simulation Results}
We illustrate the efficacy of the proposed methods in two
examples.\footnote{We have chosen on purpose low-dimensional
    examples to be able to fully detail the identified Koopman
  eigenvalues and associated subspaces. However, it is worth pointing
  out that the results presented here are applicable without any
  restriction on the type of dynamical system, its dimension, or the
  sparsity of the model in the dictionary.}

\begin{example}\longthmtitle{Unstable Discrete-time Polynomial System}
  Consider the nonlinear system
  \begin{align}\label{eq:example-unstable-polyflow}
    x_1^+ & = 1.1 \, x_1
    \nonumber \\
    x_2^+ & = 1.2 \, x_2 + 0.1 \, x_1^2 + 0.1 ,
  \end{align}
  with state $x^T = [x_1,
  x_2]^T$. System~\eqref{eq:example-unstable-polyflow} is actually an
  unstable Polyflow~\cite{RMJ-PT:19} which has a finite-dimensional
  Koopman-invariant subspace comprised of polynomials. We use the
  dictionary $D(x) = [1, x_1, x_2, x_1^2, x_1x_2, x_2^2, x_1^3, x_1
  x_2^2, x_1^2 x_2, x_2^3 ]$ with $N_d = 10$. Moreover, we gather $2
  \times 10^4$ data snapshots uniformly sampled from $[-2,2] \times
  [-2,2]$. We use the SSD and SSSD strategies to identify the maximal
  Koopman-invariant subspaces in $\Span(D(x))$. In the SSSD method, we
  use the first~10 data snapshots as signature snapshots and feed the
  rest of the data to the algorithm according to the order they appear
  in the data set. Similarly to the previous example, we use the
  strategy explained in Remark~\ref{r:finite-precision-approx} with
  $\epsilon = 10^{-12}$ to overcome error due to the use of
  finite-precision machines. Both methods find bases for the
  6-dimensional subspace spanned by $\{1, x_1, x_2, x_1x_2, x_1^2,
  x_1^3\}$, which is the maximal Koopman-invariant subspace in
  $\Span(D(x))$. The SSSD method, however, performs the calculations
  96\% faster than SSD. \new{One can find $\Kssd$ by applying EDMD on
    either of the identified dictionaries according
    to~\eqref{eq:Kssd-closed-form}}. Moreover, based on
  Theorems~\ref{t:linear-evolutions-ssd-algorithm}
  and~\ref{t:eig-identification-ssd}, we use the eigendecomposition of
  $\Kssd$ to find all the Koopman eigenfunctions associated with the
  system~\eqref{eq:example-unstable-polyflow} in
  $\Span(D(x))$. Table~\ref{table:eigenfunctions-unstable-polyflow}
  shows the identified eigenfunctions. One can use direct calculation
  to verify that the identified functions are the Koopman
  eigenfunctions associated with
  system~\eqref{eq:example-unstable-polyflow}. Note that since $x_1$
  and $x_2$ are both in the span of the identified Koopman-invariant
  subspace, one can fully characterize the behavior of the system
  using the eigenfunctions and~\eqref{eq:function-evolution-Koopman}
  linearly or directly using the identified dictionary and $\Kssd$.

  { 
    \renewcommand{\arraystretch}{1.5}
    \begin{table}[htb]
      \centering
      \caption{Identified eigenfunctions and eigenvalues of the
        Koopman operator associated with
        system~\eqref{eq:example-unstable-polyflow}.}\label{table:eigenfunctions-unstable-polyflow}
      \begin{tabular}[htb]{ | l | l | }
        \hline
        \textbf{Eigenfunction} & \textbf{Eigenvalue} \\ \hline 
        $\phi_1(x) = 1 $  & $\lambda_1 = 1$ \\ \hline 
        $\phi_2(x) = x_1 $  & $\lambda_2 = 1.1$ \\ \hline
        $\phi_3(x) = x_1^2$  & $\lambda_3= 1.21$ \\ \hline
        $\phi_4(x) = 20 \, x_1^2 - 2 \, x_2- 1$  & $\lambda_4 = 1.2$ \\ \hline
        $\phi_5(x) = x_1^3 $  & $\lambda_5 = 1.331$ \\ \hline
        $\phi_6(x) = 20 \, x_1^3 - 2 \,x_1x_2- x_1$  & $\lambda_6 = 1.32$ \\ \hline
      \end{tabular}
    \end{table}
  } 

  \new{Next, we evaluate the effectiveness of the original dictionary
    $D$ and the dictionary $\tilde{D}$ identified by SSD
    (equivalently, by SSSD) for long-term prediction.  To do this, we
    consider error functions defined as follows.  Given an arbitrary
    dictionary $\mathcal{D}$, consider its associated matrix $K =
    \EDMD {\mathcal{D}}{X}{Y}$.  For a trajectory $\{x(k)\}_{k=0}^{M}$
    of~\eqref{eq:example-unstable-polyflow} with length $M$ and
    initial condition $x_0$, let
    \begin{subequations}\label{eq:prediction-errors}
      \begin{align}
        E_{\text{relative}}(k) &= \frac{\big\| \Dc(x(k)) - \Dc(x_0)
          K^k \big\|_2}{ \| \Dc(x(k)) \|_2} \times 100
        , \label{eq:relative-prediction-error}
        \\
        E_{\text{angle}}(k) &= \angle \big(\Dc(x(k)),\Dc(x_0) K^k
        \big) , \label{eq:angle-prediction-error}
      \end{align}      
    \end{subequations}
    where $\Dc(x_0) K^k$ is the predicted dictionary vector at time
    $k$ calculated using the dictionary~$\Dc$. $ E_{\text{relative}}$
    corresponds to the relative error in magnitude between the
    predicted and exact dictionary vectors and $ E_{\text{angle}}$
    corresponds to the error in the angle of the vectors.

    We compute the errors associated to the original dictionary~$D$,
    denoted $E_{\text{relative}}^{\text{Orig}}$ and
    $E_{\text{angle}}^{\text{Orig}}$, and the errors associated to the
    SSD dictionary $\tilde{D}$, denoted
    $E_{\text{relative}}^{\text{SSD}}$ and
    $E_{\text{angle}}^{\text{SSD}}$.} 
  Figure~\ref{fig:unstable-polyflow} illustrates these errors along a
  trajectory starting from a random initial condition in $[-2,2]
  \times [-2,2]$ for 20 time steps.  \new{The plot shows the
    importance of the dictionary selection when performing
    EDMD. Unlike the prediction on $\Span(D(x))$, the prediction on
    the SSD subspace $\Span(\tilde{D}(x))$ matches the behavior of the
    system exactly. This is a direct consequence of the fact that
    $\Span(\tilde{D}(x))$ is a Koopman-invariant subspace, on which
    EDMD fully captures the behavior of the operator through~$\Kssd$.} 
\new{It is worth mentioning that based on Proposition~\ref{p:EDMD-prediction}, EDMD with dictionary $D$ also predicts the functions in  $\Span(\tilde{D})$ exactly. However, its prediction for functions outside of $\Span(\tilde{D})$ leads to large errors.}

  \begin{figure}[htb]
    \centering 
    {\includegraphics[width=.47
      \linewidth]{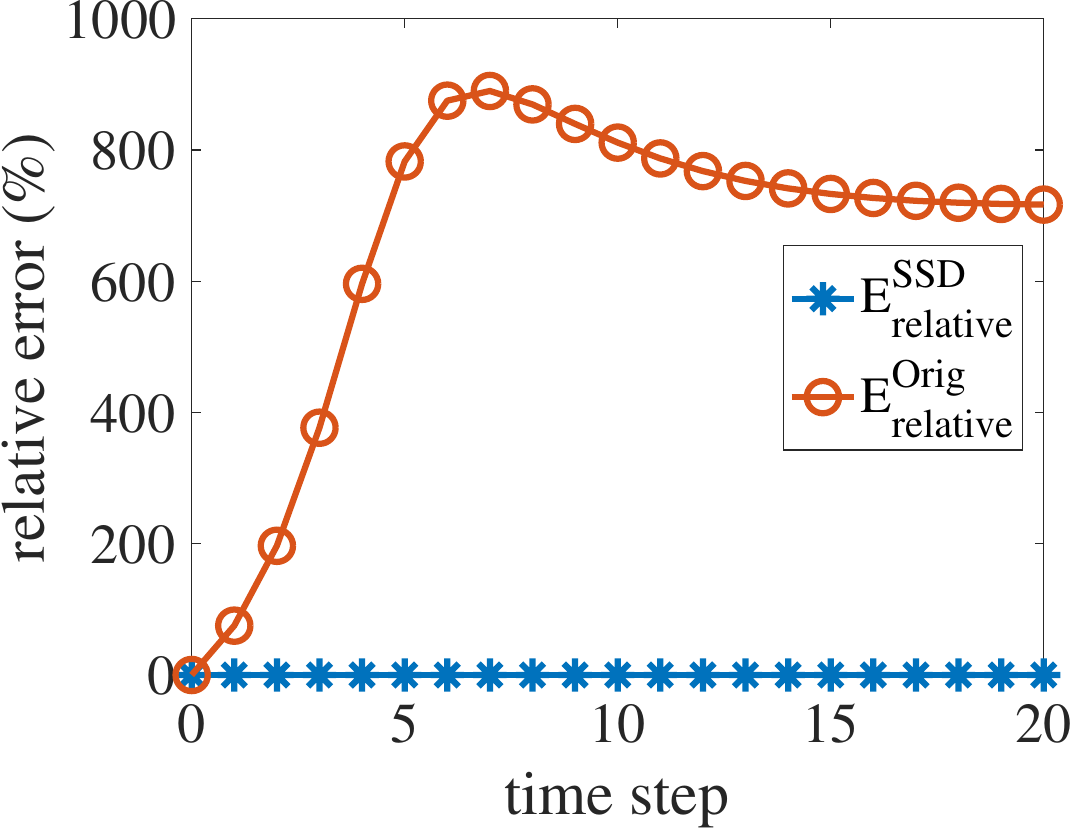}}
    {\includegraphics[width=.47
      \linewidth]{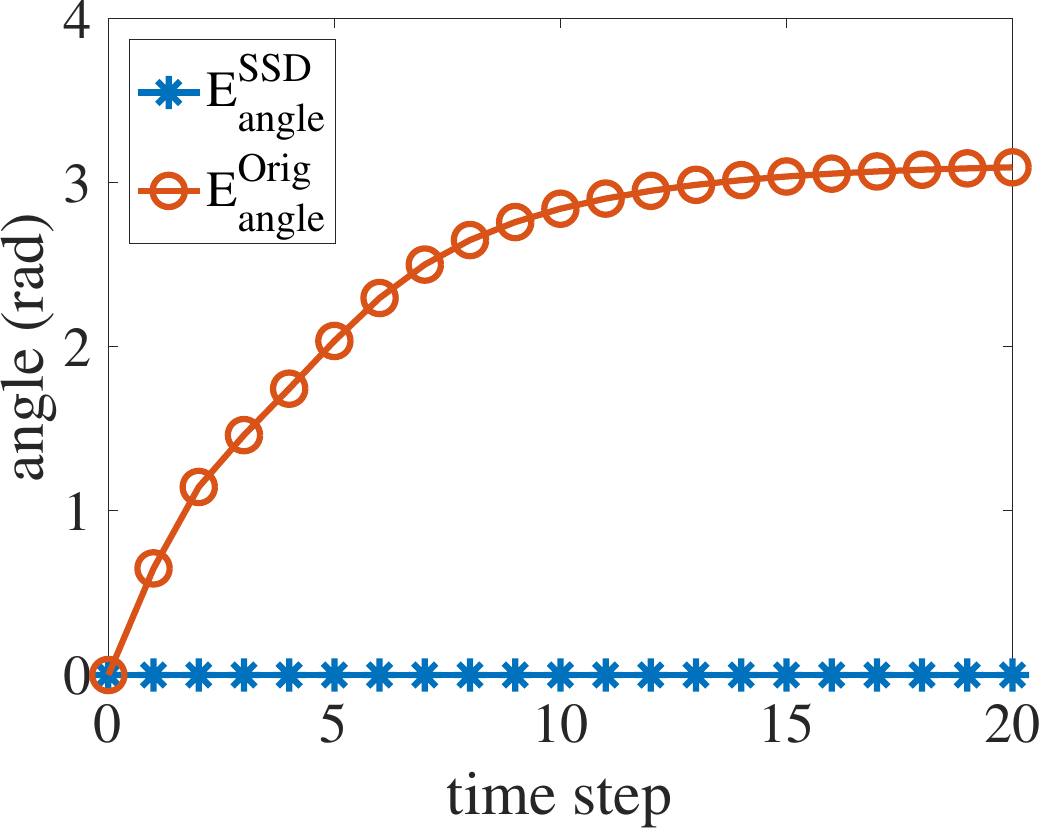}}
    \caption{\color{black}Relative (left) and angle (right) prediction errors on the original and SSD subspaces for system~\eqref{eq:example-unstable-polyflow} on
      a trajectory of length $M=20$.}\label{fig:unstable-polyflow}
\vspace*{-1ex}
  \end{figure}
\end{example}

\begin{figure}[htb]
\centering 
{\includegraphics[width=.47
\linewidth]{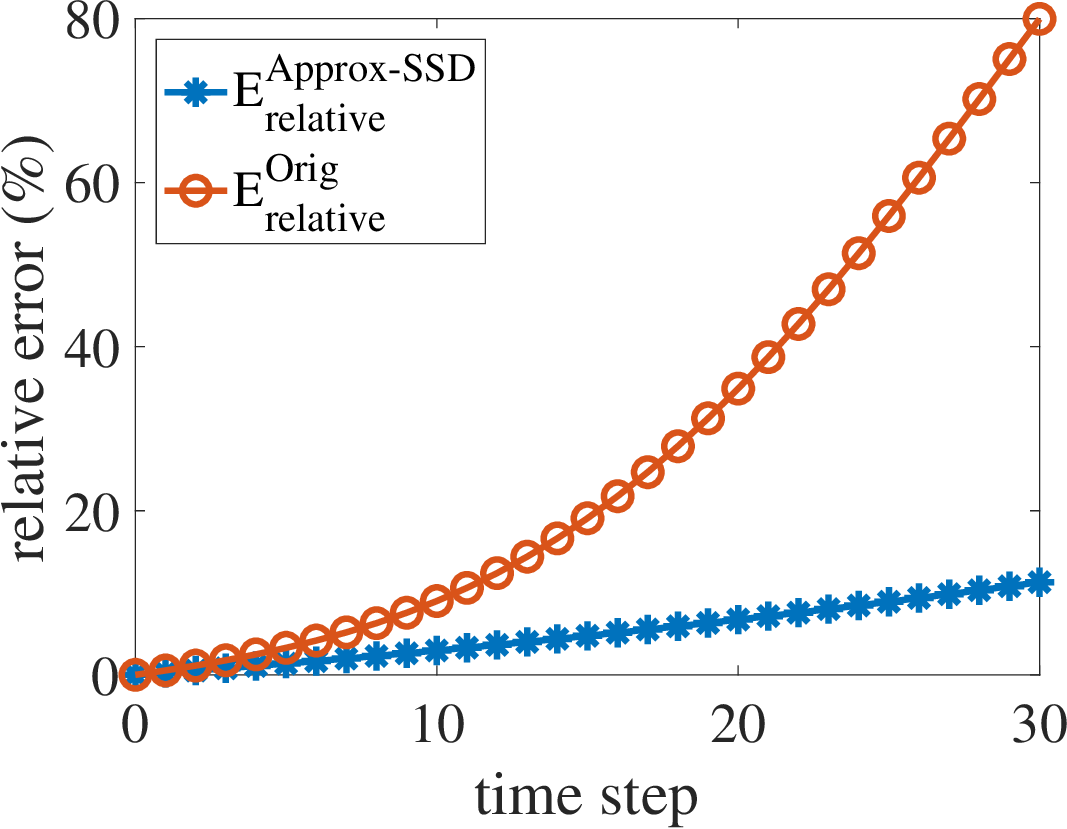}}
{\includegraphics[width=.47
\linewidth]{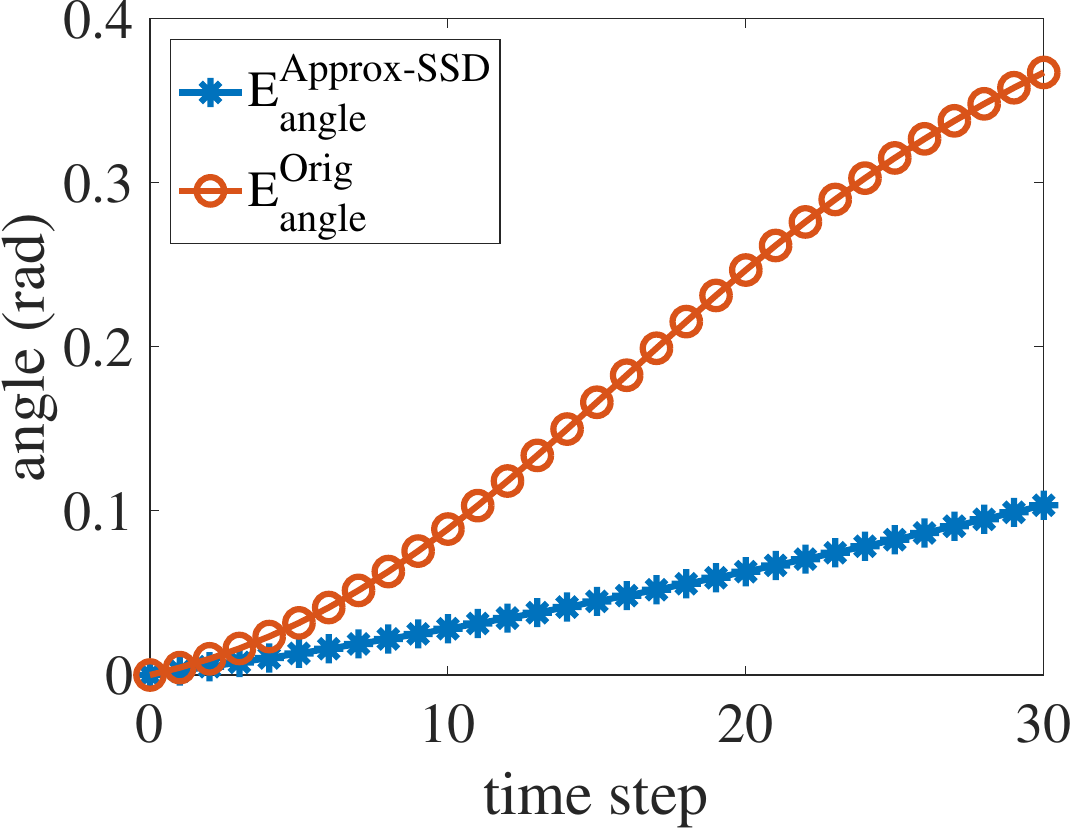}}
\caption{\color{black}{Relative (left) and angle (right) prediction
    errors on Approximated-SSD and original subspaces for system~\eqref{eq:Duffing}
    on a trajectory of length $M=30$.}}\label{fig:Duffing-prediction}
\vspace*{-2ex}
\end{figure}

\begin{example}\longthmtitle{Duffing Equation}
  Here, we investigate the efficacy of the proposed methods in
  approximating Koopman eigenfunctions and invariant subspaces.
  Consider the Duffing equation~\cite[Section 4.2]{MOW-IGK-CWR:15}
\begin{align}\label{eq:Duffing}
\dot{x}_1 &= x_2
\nonumber \\
\dot{x}_2 &=  -0.5x_2 +x_1 - x_1^3,
\end{align}
with state $x^T = [x_1, x_2]^T$. The system has one unstable
equilibrium at the origin and two locally stable equilibria at $[\pm
1,0]^T$. We consider the discretized version of~\eqref{eq:Duffing}
with timestep $\Delta t = 0.01s$ and gather $N = 5000$ data snapshots
uniformly sampled from $\Mc = [-2,2] \times [-2,2]$. Moreover, we use
the dictionary $D$ comprised of all $N_d = 36$ monomials up to degree
7 in the form of $\prod_{i=1}^{7} y_i$, where $y_i \in
\{1,x_1,x_2\}$. The maximal Koopman-invariant subspace in the span of
the dictionary is one dimensional, spanned by the trivial
eigenfunction $\phi(x) \equiv 1$. Hence, applying the SSD and SSSD
algorithms would result in a trivial solution. Instead, we apply the
Approximated-SSD algorithm on the available dictionary snapshots with
the accuracy parameter $\epsilon = 10^{-3}$. The outcome is the
dictionary $\Dtilde$ with $\tilde{N}_d = 15$ elements. We calculate
the linear prediction matrix $\Kssdapprox$ using
Theorem~\ref{t:approximated-ssd-accuracy}. The norm of the
perturbation $\| [\Delta_1^*, \Delta_2^*] \|_F $ satisfies
\begin{align*}
  \| [\Delta_1^*, \Delta_2^*] \|_F \approx 9.6 \times 10^{-4} \|
  [\Dtilde(X), \Dtilde(Y)] \|_F,
\end{align*}
agreeing with the upper bound provided in
Theorem~\ref{t:approximated-ssd-accuracy}. We approximate the
eigenfunctions of the Koopman operator using the eigendecomposition of
$\Kssdapprox$. For space reasons, we only illustrate the leading
nontrivial approximated Koopman eigenfunctions with eigenvalue closest
to the unit circle. Figure~\ref{fig:Duffing-real-eigenfunction}(left)
shows the real-valued approximated eigenfunction corresponding to the
eigenvalue $\lambda = 0.9919$. Despite being an approximation, the
eigenfunction captures the behavior of the vector field accurately and
correctly identifies the attractiveness of the two locally stable
equilibria. Given that $|\lambda | < 1$,
Figure~\ref{fig:Duffing-real-eigenfunction}(left) predicts that the
trajectories eventually converge to one of the stable
equilibria. Figure~\ref{fig:Duffing-real-eigenfunction}(right) shows
the absolute value of the approximated Koopman eigenfunctions
corresponding to the eigenvalues $\lambda = 0.9989 \pm 0.0037j
$. Similarly to the other plot, it captures information about the
shape of the vector field such as the attractive equilibria and their
regions of attraction.
\begin{figure}[htb]
  \centering
  {\includegraphics[width=.47\linewidth]{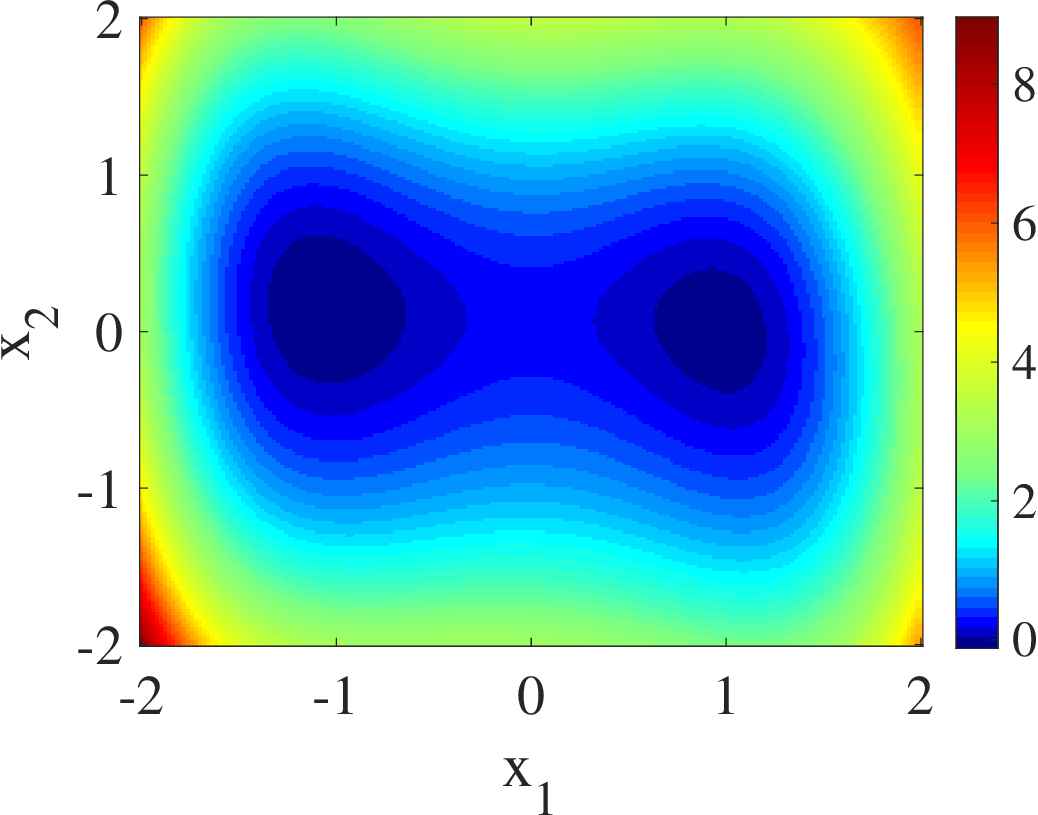}}
  {\includegraphics[width=.49\linewidth]{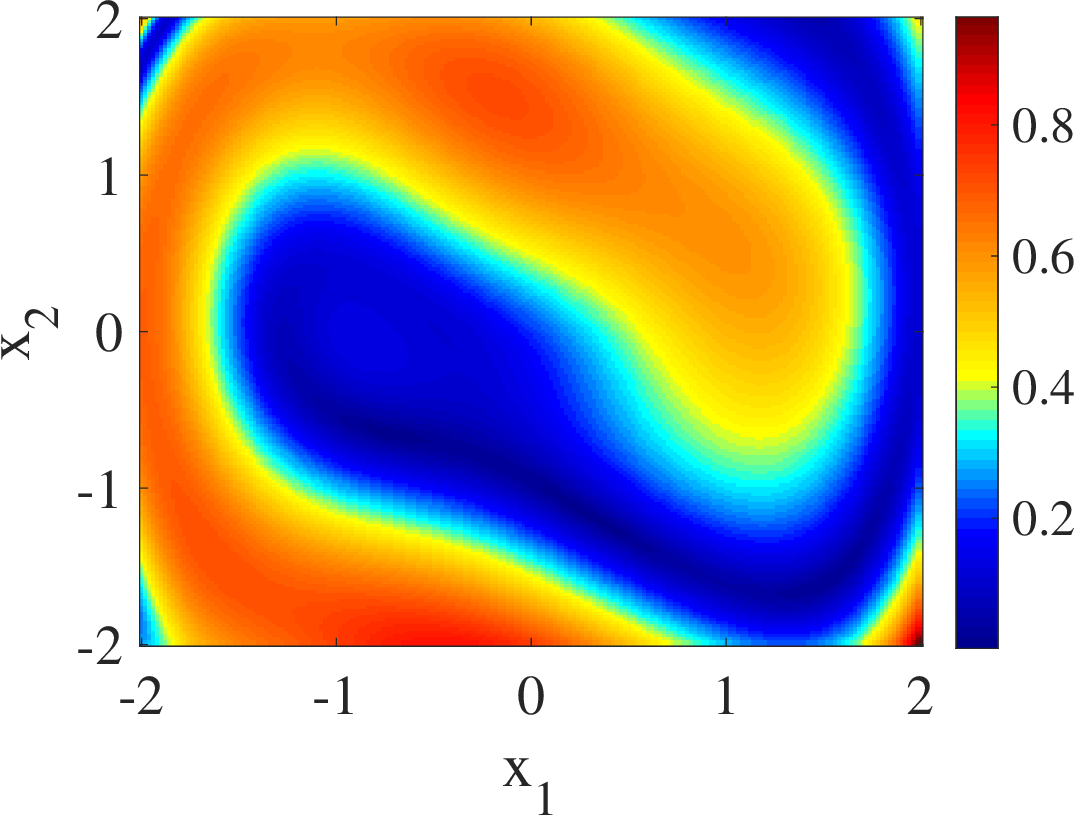}}
  \caption{\color{black}{The approximated eigenfunction corresponding
      to the eigenvalue $\lambda = 0.9919$ (left) and the absolute
      value of the approximated eigenfunctions corresponding to the
      eigenvalues $\lambda = 0.9989 \pm 0.0037j $ (right) for the
      Koopman operator associated with~\eqref{eq:Duffing}, as
      identified by the Approximated-SSD
      algorithm.}}\label{fig:Duffing-real-eigenfunction}
\vspace*{-2ex}
\end{figure}

\new{We use the relative and angle errors defined
  in~\eqref{eq:prediction-errors} to compare the prediction accuracy
  of the original dictionary $D$ and the Approximated-SSD dictionary
  $\Dtilde$ (for which we use $\Kssdapprox$).
  Figure~\ref{fig:Duffing-prediction} illustrates the relative and
  angle errors along a trajectory starting from a random initial
  condition in $[-2,2] \times [-2,2]$ for 30 timesteps. The plot shows
  the superiority of EDMD over the subspace identified by
  Approximated-SSD in long-term prediction of dynamical behavior.  }
\end{example}

\section{Conclusions}
We have studied the characterization of Koopman-invariant subspaces
and Koopman eigenfunctions associated to a dynamical system by means
of data-driven methods.  We have shown that the application of EDMD
over a given dictionary forward and backward in time fully
characterizes whether a function evolves linearly in time according to
the available data.  Building on this result, and under dense
sampling, we have established that functions satisfying this condition
are Koopman eigenfunctions almost surely.
We have developed the SSD algorithm to identify the maximal
Koopman-invariant subspace in the span of the given dictionary and
formally characterized its correctness.
Finally, we have developed extensions to scenarios with large and
streaming data sets, where the algorithm refines its output as new
data becomes available,
and to scenarios where the original dictionary does not contain
sufficient informative eigenfunctions, in which case the algorithm
obtains instead approximations of the Koopman eigenfunctions and
invariant subspaces.  Future work will develop parallel and
distributed counterparts of the algorithms proposed here over network
systems, \new{obtain out-of-sample bounds on prediction accuracy for
  the output of the Approximated-SSD algorithm,} investigate the
design of noise-resilient methods to identify Koopman eigenfunctions
and invariant subspaces, and explore methods to expand the original
dictionary to ensure the existence of non-trivial invariant subspaces.

\bibliographystyle{IEEEtran}%
\bibliography{alias,JC,Main,Main-add}

\appendix

Here we gather some linear algebraic results. 

\begin{lemma}\longthmtitle{Intersection of Linear
    Spaces}\label{l:subspace-intersection}
  Let $A, B \in \real^{m \times n}$ be matrices with full column rank.
  Suppose that the columns of
  $Z=[(Z^A)^T,(Z^B)^T]^T \in \real^{2n \times l}$ form a basis for the
  null space of $[A,B]$, where $Z^A,Z^B \in \real^{n \times l}$. Then,
  \begin{enumerate}
  \item $\range(AZ^A) = \range(A) \cap \range(B)$;
  \item $Z^A$ and $Z^B$ have full column rank.
  \end{enumerate}
\end{lemma}
\begin{proof}
  (a) First, note that $\range(AZ^A) \subseteq \range(A)$. Moreover,
  by hypothesis, $[A,B]Z=0$, which leads to
  $\range(AZ^A) = \range(BZ^B) \subseteq \range(B)$. Consequently,
  $\range(AZ^A) \subseteq \range(A) \cap \range(B)$.
  Now, suppose that $ v \in \range(A) \cap \range(B)$. By
  definition, there
  exist vectors $w_1,w_2 \in \real ^n$ such that $v=A w_1=B
  w_2$. Then,
  \begin{align*}
    \begin{bmatrix}
      A, B
    \end{bmatrix}
    \begin{bmatrix}
      w_1 
      \\
      -w_2 
    \end{bmatrix}
    = 0,
  \end{align*}
  which means that $[w_1^T,-w_2^T]^T \in \range(Z)$ and there exists
  $r \in \real ^l$ such that $w_1 = Z^A r$ and $w_2 = - Z^B
  r$. Therefore, $ v = Aw_1 = AZ^A r \in
  \range(AZ^A)$. Consequently, $\range(A) \cap \range(B) \subseteq \range(AZ^A)$
  and the claim follows.
  
  (b) We prove this part using contradiction. Suppose that there
  exists $v\neq 0$ such that $Z^A v =0$. Also, since $[A,B]Z =0$, one
  can conclude that $A Z^A v = - B Z^B v =0$. Since $B$ has full
  column rank, we have $Z^B v =0$. Hence $Zv =0$, which is a
  contradiction since the columns of $Z$ are linearly independent. A
  similar reasoning shows that $Z^B$ has full column rank.
\end{proof}

\begin{lemma}\label{l:product-subspace}
  Let $A,C,D$ be matrices of appropriate sizes, with $A$ having full
  column rank. Then $\range(AC) \subseteq \range(AD)$ if and only if
  $\range(C) \subseteq \range(D)$.
\end{lemma}
\begin{proof}
  $(\Rightarrow)$: Suppose that $v \in \range(C)$, and hence there
  exists $w$ such that $v=Cw$. Therefore, $Av = ACw \in
  \range(AC)$. Since $\range(AC) \subseteq \range(AD)$, one can deduce
  that there exist $r$ such that $ACw=ADr$, and we get
  $A(v-Dr)=0$. This leads to $v=Dr \in \range(D)$ since $A$ has full
  column rank, and hence $\range(C) \subseteq \range(D)$.
  
  $(\Leftarrow)$: Suppose that $v \in \range(AC)$, and hence $v=ACw$
  for some $w$. Since $\range(C) \subseteq \range(D)$, there exists
  $r$ such that $Cw=Dr$. As a result, $v =ACw=ADr$ which leads to the
  conclusion that $v \in \range(AD)$ and the claim follows.
\end{proof}

\begin{lemma}\label{l:symmetric-immersion}
  Let $A_1,B_1 \in \real^{m \times n}$, $A_2,B_2 \in \real^{l \times
    n}$, and $C \in \real^{n \times k}$ with $A_1, B_1, C$ having full
  column rank. If
   \begin{subequations}
     \begin{align}
       \range\Bigg(\begin{bmatrix} A_1 \\ A_2 \end{bmatrix} \Bigg) &=
       \range\Bigg(\begin{bmatrix} B_1 \\ B_2 \end{bmatrix}
       \Bigg), \quad \label{eq:si-1}
       \\
       \range(A_1 C) &= \range (B_1 C),   \label{eq:si-2}
     \end{align}
   \end{subequations}
  then
  \begin{align*}
    \range\Bigg(\begin{bmatrix} A_1 \\ A_2 \end{bmatrix} C \Bigg) &=
    \range\Bigg(\begin{bmatrix} B_1 \\ B_2 \end{bmatrix} C \Bigg).
  \end{align*}
\end{lemma}
\begin{proof}
  Based on~\eqref{eq:si-1}, one can deduce that there exists a
  nonsingular square matrix $K$ such that
  \begin{subequations}
    \begin{align}
      A_1 = B_1 K, \label{eq:si-3}
      \\
      A_2 = B_2 K.\label{eq:si-4}
    \end{align}
  \end{subequations}
  Multiplying both sides of~\eqref{eq:si-3} by $C$ gives
  \begin{align}\label{eq:si-5}
    A_1 C = B_1 K C.
  \end{align}
  Moreover, based on~\eqref{eq:si-2}, one can deduce that there exists
  a nonsingular square matrix $K^*$
  \begin{align}\label{eq:si}
    A_1 C = B_1 C K^*.
  \end{align}
  By subtracting~\eqref{eq:si} from~\eqref{eq:si-5} and considering
  the fact that 
   $B_1$ has full column rank,
  one can deduce
  that
  \begin{align}\label{eq:si-6}
    C K^* = K C.
  \end{align}
  Now, multiplying both sides of~\eqref{eq:si-4} from the right by $C$
  and using~\eqref{eq:si-6}, one can write $A_2 C = B_2 C K^*$, which
  in conjunction with~\eqref{eq:si} leads to
  \begin{align*}
    \begin{bmatrix} A_1 \\ A_2 \end{bmatrix} C = \begin{bmatrix} B_1
      \\ B_2 \end{bmatrix} C K^*,
  \end{align*}
   completing the proof.
\end{proof}

\begin{lemma}\label{l:infinite-intersection-product}
  Let $A$, $\{C_i\}_{i =1}^\infty$, and $\hat{C}$ be matrices of
  appropriate sizes. Assume that $A$ has full column rank and
  $\range(\hat{C}) = \bigcap_{i=1}^\infty \range(C_i)$. Then
  $ \range(A \hat{C})=\bigcap_{i=1}^\infty \range(A C_i)$.
\end{lemma}
\begin{proof}
  First, we prove that
  $\range(A \hat{C}) \subseteq \bigcap_{i=1}^\infty \range(A
  C_i)$. Let $v \in \range (A \hat{C})$, then there exists a vector
  $w$ such that $v = A \hat{C} w = A r$, with $r = \hat{C} w$. Note
  that $r \in \range(\hat{C})$ and consequently $r \in \range(C_i)$
  for all $i \in \naturals$. Hence, for every $i \in \naturals$ there
  exists $z_i$ such that $r = C_i z_i$.  Based on the definition of
  $v$, we have $v = A r = A C_i z_i$ for every $i \in \naturals$. As a
  result, $v \in \bigcap_{i=1}^\infty \range(A C_i)$ which concludes
  the proof of this inclusion.
  
  Now, we prove that
  $\bigcap_{i=1}^\infty \range(A C_i) \subseteq \range(A
  \hat{C})$. Let $v \in \bigcap_{i=1}^\infty \range(A C_i)$. Then for
  every $i \in \naturals$, there exists $w_i$ such that
  $v = A C_i w_i$. Moreover, $ v \in \range(A)$ and since $A$ has full
  column rank, there exists a unique $r$ such that $v =
  Ar$. Therefore, for all $i \in \naturals$, we have $r = C_i
  w_i$. Thus, $r \in \bigcap_{i=1}^\infty \range(C_i)$ and
  consequently, $r \in \range(\hat{C})$. Since $v = Ar$, we have
  $v \in \range(A \hat{C})$, concluding the proof.
\end{proof}

\vspace*{-3ex}

\begin{IEEEbiography}[{\includegraphics[width=1in,height=1.25in,clip,keepaspectratio]{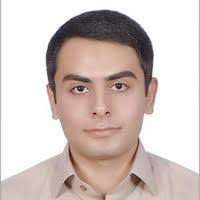}}]{Masih
    Haseli} was born in Kermanshah, Iran in 1991. He received the
  B.Sc. degree, in 2013, and M.Sc. degree, in 2015, both in Electrical
  Engineering from Amirkabir University of Technology (Tehran
  Polytechnic), Tehran, Iran. In 2017, he joined the University of
  California, San Diego to pursue the Ph.D. degree in Mechanical and
  Aerospace Engineering.  His research interests include system
  identification, nonlinear systems, network systems, data-driven
  modeling and control, and distributed and parallel computing.
  Mr. Haseli was the recipient of the bronze medal in Iran's national
  mathematics competition in 2014.
\end{IEEEbiography}

\vspace*{-4ex}

\begin{IEEEbiography}[{\includegraphics[width=1in,height=1.25in,clip,keepaspectratio]{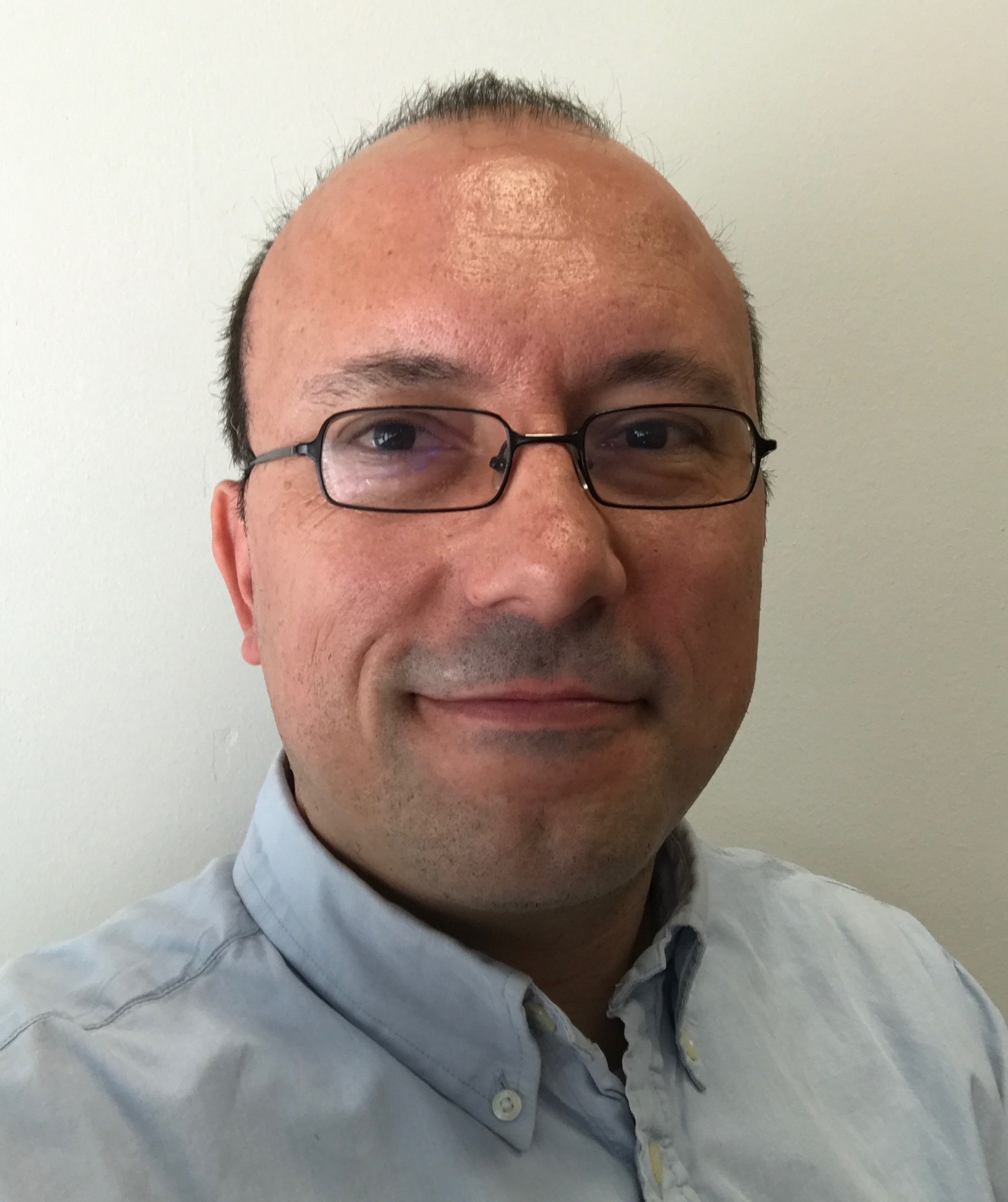}}]{Jorge
    Cort\'{e}s}
  (M'02, SM'06, F'14) received the Licenciatura degree in mathematics
  from Universidad de Zaragoza, Zaragoza, Spain, in 1997, and the
  Ph.D. degree in engineering mathematics from Universidad Carlos III
  de Madrid, Madrid, Spain, in 2001. He held postdoctoral positions
  with the University of Twente, Twente, The Netherlands, and the
  University of Illinois at Urbana-Champaign, Urbana, IL, USA. He was
  an Assistant Professor with the Department of Applied Mathematics
  and Statistics, University of California, Santa Cruz, CA, USA, from
  2004 to 2007. He is currently a Professor in the Department of
  Mechanical and Aerospace Engineering, University of California, San
  Diego, CA, USA. He is the author of Geometric, Control and Numerical
  Aspects of Nonholonomic Systems (Springer-Verlag, 2002) and
  co-author (together with F. Bullo and S.  Mart{\'\i}nez) of
  Distributed Control of Robotic Networks (Princeton University Press,
  2009).  He is a Fellow of IEEE and
  SIAM. 
  His current research interests include distributed control and
  optimization, network science, resource-aware control, nonsmooth
  analysis, reasoning and decision making under uncertainty, network
  neuroscience, and multi-agent coordination in robotic, power, and
  transportation networks.
\end{IEEEbiography}

\end{document}